\documentclass[aps,longbibliography,showpacs,twocolumn,superscriptaddress,amsmath,amssymb,verbatim]{revtex4-2}
\usepackage[thinlines]{easytable}
\usepackage{graphicx}
\usepackage{lmodern}
\usepackage{epstopdf}
\usepackage{dcolumn}
\usepackage{bm}
\usepackage{subfigure}
\usepackage{makecell}
\usepackage{amsmath} 
\usepackage[pdftex,colorlinks=true,citecolor=blue,linkcolor=blue,urlcolor=blue,bookmarks=true]{hyperref}
\usepackage{booktabs} 
\usepackage{amsmath}  

\usepackage{color}
\usepackage{textcomp}
\definecolor{red}{rgb}{1,0,0}

\definecolor{blue}{rgb}{0,0,1}

\definecolor{green}{rgb}{0,1,0}

\begin{document}
\title{Spin dynamics and 1/3 magnetization plateau in a coupled distorted diamond  chain compound K$_2$Cu$_3$(MoO$_4$)$_4$ \\ }
\author{G. Senthil Murugan\textsuperscript{†}}
\homepage{nanosen@gmail.com}
\affiliation{Institute of Physics, Academia Sinica, Taipei 11529, Taiwan}	
\affiliation{Department of Physics, St. Joseph's College of Engineering, OMR, Chennai 600 119, India}
\author{J. Khatua}
\thanks{These authors contributed equally to this work}
\affiliation{Department of Physics, Sungkyunkwan University, Suwon 16419, Republic of Korea}
\author{Suyoung Kim}
\affiliation{Department of Physics, Simon Fraser University, Burnaby, BC V5A 1S6, Canada}
\author{Eundeok Mun}
\affiliation{Department of Physics, Simon Fraser University, Burnaby, BC V5A 1S6, Canada}
  \author{K.  Ramesh Babu}
  \affiliation{Department of Physics, Anurag Engineering College (An UGC Autonomous Institution), Ananthagiri (V $\&$ M), Kodad, Suryapet (D), Telangana 508206, India}
  \author{Heung-Sik Kim}
  \affiliation{
  	Department of Semiconductor Physics and Institute of Quantum Convergence Technology, Kangwon National University, Chuncheon 24341, Republic of
  	Korea}
  \author{C.-L. Huang} 
   \affiliation{Department of Physics, National Cheng Kung University, Tainan 701, Taiwan}
   \author{R. Kalaivanan}
   \affiliation{Institute of Physics, Academia Sinica, Taipei 11529, Taiwan}
  \author{U. Rajesh Kumar}  
  \affiliation{Institute of Physics, Academia Sinica, Taipei 11529, Taiwan}
  \affiliation{Centre for Nanotecchnology, Indian Institute of Technology Roorkee, 247667, India}
  \author{I. Panneer Muthuselvam}
  \affiliation{Department of Physics (MMV), Banaras Hindu University, Varanasi 221005, Uttar Pradesh, India}
  \author{W. T. Chen} 
  \affiliation{Center for Condensed Matter Sciences, National Taiwan University, Taipei 10617, Taiwan}
  \author{Sritharan Krishnamoorthi}
  \affiliation{Institute of Physics, Academia Sinica, Taipei 11529, Taiwan}
   \author{K.-Y. Choi}
   \homepage{choisky99@skku.edu}
   \affiliation{Department of Physics, Sungkyunkwan University, Suwon 16419, Republic of Korea}
     \author{R. Sankar}
     \homepage{sankarndf@gmail.com}
     \affiliation{Institute of Physics, Academia Sinica, Taipei 11529, Taiwan}
    
\date{\today}

\begin{abstract}
We investigate magnetic properties of the $s$ = 1/2 compound K$_{2}$Cu$_{3}$(MoO$_{4}$)$_{4}$ by combining magnetic susceptibility, magnetization, specific heat, and electron spin resonance (ESR) with density functional calculations. Its monoclinic structure features alternating Cu$^{2+}$ ($s$ = 1/2) monomers and edge-shared dimers linked by MoO$_{4}$ units, forming a distorted diamond chain along the $a$-axis. Antiferromagnetic order occurs at $T_{\rm N}$ = 2.3 K, as evident from a $\lambda$-type anomaly in specific heat and magnetic susceptibility derivatives. Inverse magnetic susceptibility  reveals coexisting ferro- and antiferromagnetic interactions. Specific heat and ESR data show two characteristic temperatures: one at 20 K, associated with spin-singlet formation in Cu$_{2}$O$_{9}$ dimers, and another at 3.68 K, indicating short-range correlations between dimers and monomers. Magnetization measurements reveal a metamagnetic transition at 2.6 T and a critical magnetic field $\mu_{0}H_{c}$ = 3.4 T, where a 1/3 magnetization plateau emerges with saturation near 0.35 $\mu_{\rm B}$. Low-temperature specific heat and magnetization data reveal the  suppression of long-range order at $\mu_{0}H_{c}$, enabling the construction of a temperature-magnetic field phase diagram showing multiple magnetic phases near the $\mu_{0}H_{c}$. Density functional theory confirms a distorted diamond chain with $J_{1}$ dimers and competing $J_2$, $J_4$, $J_3$, and $J_5$ interactions with monomer spins as an effective low-temperature spin model.
\end{abstract}

\maketitle

\section{\label{sec:level1} Introduction\protect\\ }

Frustration in quantum magnets arises from either the arrangement of magnetic spins on lattices (geometric frustration) or the presence of competing exchange interactions (exchange frustration). In both scenarios, the persistent magnetic interactions fail to collectively minimize their energies, resulting in fluctuations even at $T$ = 0 K. This frustration yields a system with a degenerate ground state and a notable suppression of long-range magnetic ordering, giving rise to a variety of exotic magnetic phases such as valence bond solids, spin nematics, spin liquids, and spin ice~\cite{ANDERSON1973153,Matan2010,Balents2010,RevModPhys.89.025003,Savary_2016,doi:10.1126/science.aay0668,Jeon2024,KHATUA20231}.\\The diamond spin chain, where diamond-shaped units form a one-dimensional (1D) lattice, is one of the simplest frustrated lattice system. Azurite, Cu$_3$(CO$_3$)$_2$(OH)$_2$, is a natural mineral featuring Cu$^{2+}$ ($s$ = 1/2) ions arranged in an alternating monomer-dimer pattern along the crystallographic $b$-axis to create an infinite chain. This structure has been originally proposed as an experimental realization of the 1D frustrating distorted diamond chain model~\cite{KIKUCHI2004900,PhysRevLett.94.227201}. \\Early studies of azurite revealed antiferromagnetic (AFM) order at approximately 1.86 K ~\cite{PhysRev.112.1544,10.1063/1.1730553,VANDERLUGT19591313,FRIKKEE1962269,LOVE1970290}. However, Kikuchi {\it et al.}~\cite{KIKUCHI2004900,PhysRevLett.94.227201} introduced fresh insights by uncovering a two-stage process in the development of magnetic short-range correlations, observed around 20 and 5 K, evident in both magnetic susceptibility and specific heat measurements. Additionally, they identified a magnetic plateau occurring at one-third of the saturation magnetization in high-field magnetization measurements at $T$ = 1.5 K. They attributed these behaviors to a triangular unit where all three interactions (intra-dimer interaction 
$J_{2}$ and two disparate interactions $J_{1}$ and $J_{3}$ between monomer and dimers)
are AFM. These findings have sparked renewed interest in the magnetism of azurite~\cite{PhysRevLett.94.227201}. \\
Despite extensive experimental and theoretical investigations \cite{PhysRevLett.97.089701,PhysRevLett.100.117202,PhysRevB.77.054405,PhysRevB.81.140406,PhysRevLett.106.217201}, it remains elusive to fully comprehend the experimental observations of double-peak appearance, magnetic plateaus, long-range AFM order, and exchange coupling strengths in azurite. However, it is reasonable to envision azurite as a two-dimensional spin lattice, where diamond chains are connected through weakly interchain exchange interactions, denoted as $J_4$~\cite{Kang_2009,PhysRevLett.102.127205,PhysRevB.84.184419,molecules26030531}.\\
Interestingly, other diamond chain systems do not exhibit magnetic properties similar to azurite \cite{ishii2000gapped,doi:10.1143/JPSJ.80.104710}. However, a recent investigation of the highly 1D spin-1/2 inequilateral diamond chain compound K$_3$Cu$_3$AlO$_2$(SO$_4$)$_4$ has revealed a double-peak pattern in magnetic susceptibility at temperatures around 50 and 200 K, approximately an order of magnitude higher than those observed in azurite \cite{doi:10.7566/JPSJ.84.073702}. Despite this enhanced magnetic energy scale, no magnetic order is observed down to 0.5 K, suggesting that K$_3$Cu$_3$AlO$_2$(SO$_4$)$_4$ represents the first material exhibiting an alternate dimer-monomer spin-liquid ground state \cite{doi:10.7566/JPSJ.84.073702}.\\
To deepen our understanding of diamond chain systems, it is crucial to synthesize a broader range of candidate systems. Herein, we report on synthesis and structural and magnetic properties of the brand-new distorted diamond chain compound K$_2$Cu$_3$(MoO$_4$)$_4$ by combining a synchrotron X-ray powder diffraction (SXRD), magnetic susceptibility, isothermal magnetization, specific heat, and ESR with density functional theory (DFT) calculations.  Magnetic susceptibility, specific heat, and magnetization measurements revealed long-range AFM order at 
$T_{\rm N}$ = 2.3 K, with a high-temperature broad maximum (20 K) linked to spin-singlet correlations and a low-temperature broad maximum (3.8 K) pertaining to short-range correlations. Isothermal magnetization data showed a metamagnetic transition at 2.6 T, followed by a field-induced transition to a 1/3 magnetization plateau at 3.4 T, typical of frustrated diamond chains. DFT calculations indicated a magnetic energy hierarchy between monomer and dimer subsystems and that AFM order is facilitated by alternating monomer spin directions along the distorted chain.
\section{Experimental and DFT calculation details}  Polycrystalline samples of K$_2$Cu$_3$(MoO$_4$)$_4$ were prepared by a solid-state reaction method using the initial ingredients of K$_2$CO$_3$, CuO, and MoO$_3$ (higher than $99.95\%$ purity). Prior to use the reagents were preheated in air at 100$^\circ$C for 24 hrs to eliminate moisture. The mixture was then sintered at 400$^\circ$C and 500$^\circ$C in air for 48 hrs each with intermediate grindings. Finally, the sample was annealed at 550$^\circ$C for 72 hrs to get the single phase of K$_2$Cu$_3$(MoO$_4$)$_4$. The phase purity of the obtained polycrystalline sample was checked by SXRD technique at room temperature  using the MYTHEN detector with 15 keV beam at the 09A beamline of NSRRC in Taiwan. The collected pattern was analyzed with the Rietveld method using the software of Bruker TOPAS. \\Magnetic measurements were carried out  using a superconducting quantum interference device vibrating-sample magnetometer (SQUID-VSM, Quantum Design, USA) in the temperature range 2 K $\leq$ $T$ $\leq$ 300 K and in magnetic fields up to 7 T. Furthermore, low-temperature
magnetization measurements were carried out
 down to 0.4 K using the $^{3}$He option of the MPMS3 SQUID magnetometer
from Quantum Design. Specific heat measurements were performed using a standard relaxation method with a physical property measurement system (PPMS, Quantum Design, USA) in the temperature range 2 K $\leq$ $T$ $\leq$ 300 K and in magnetic fields up to 9 T. In addition,  specific heat was measured in the temperature range 0.3 K $\leq$ $T$ $\leq$ 3.8 K using a Quantum Design Dynacool PPMS equipped with $^{3}$He option in several magnetic fields up to 9 T.
\\ X-band measurements were performed on polycrystalline samples of K$_{2}$Cu$_{3}$(MoO$_{4}$)$_{4}$  at a frequency of $\nu$ = 9.5
GHz. A Bruker EMXplus-9.5/12/P/L spectrometer was employed with a continuous He flow cryostat to carry out ESR measurements over the temperature
range of 5 K $\leq$ $T$ $\leq$ 295 K.\\
Considering the complexities of exchange interactions in  diamond chain compounds, we performed two separate DFT calculations. In order to give an appropriate description of the electron-electron correlations associated with the 3$d$ electrons of the Cu atom, we first applied the generalized gradient approximation (GGA $+$ $U_{\rm eff}$) method  with $U_{\rm eff}$ = $U$ – $J_{H}$ = 3.6 eV, where $U$ represents the on-site coulomb repulsion and $J_{H}$ denotes the Hund's coupling \cite{PhysRevLett.77.3865,PhysRevB.57.1505,Petersen2006}. We used the projector augmented-wave method as implemented in the Vienna ab initio simulation package (VASP) \cite{PhysRevB.59.1758,PhysRevB.47.558,PhysRevB.49.14251}. Wavefunctions were expanded in a plane-wave basis with energy cut-off of 400 eV. The Brillouin zone integration was performed with a $\Gamma$-centered Monkhorst-Pack k-point mesh of 2 $\times$ 6 $\times$ 8. The self-consistent total energies were converged up to 10$^{-4}$ eV.  This first method, whose results are presented in Table~\ref{listJ}, determines exchange interactions by comparing the total energy differences of five distinct magnetic configurations, as shown in Fig.~\ref{fig-SS}. However, this approach has limitations in accounting for longer-range magnetic exchange interactions, as it considers only the exchange interactions depicted in Fig.~\ref{LDA} and is restricted to the $\Gamma$-point of the Brillouin zone.\\ To complement this, we employed the magnetic-force linear response theory \cite{LIECHTENSTEIN198765}, which inherently incorporates long-range magnetic exchange interactions and accounts for the effects of $q\neq 0$. 
These DFT calculations were conducted using the local density approximation (LDA) as the exchange-correlation functional for different values of $U$.  The corresponding DFT + $U$ (LDA) calculations were performed using the openMX
software package, which employs DFT and a linear combination of pseudoatomic orbitals \cite{PhysRevB.67.155108}. The result of the  self-consistent field calculation obtained from openMX was used as input for the $J_{X}$ code \cite{YOON2020106927}, which calculates the exchange coupling parameters $J_{ij}^{\rm LDA}$ between localized spins using the Green’s function representation of the Liechtenstein formula.
\begin{figure*}
	\centering
	\includegraphics[width=\textwidth]{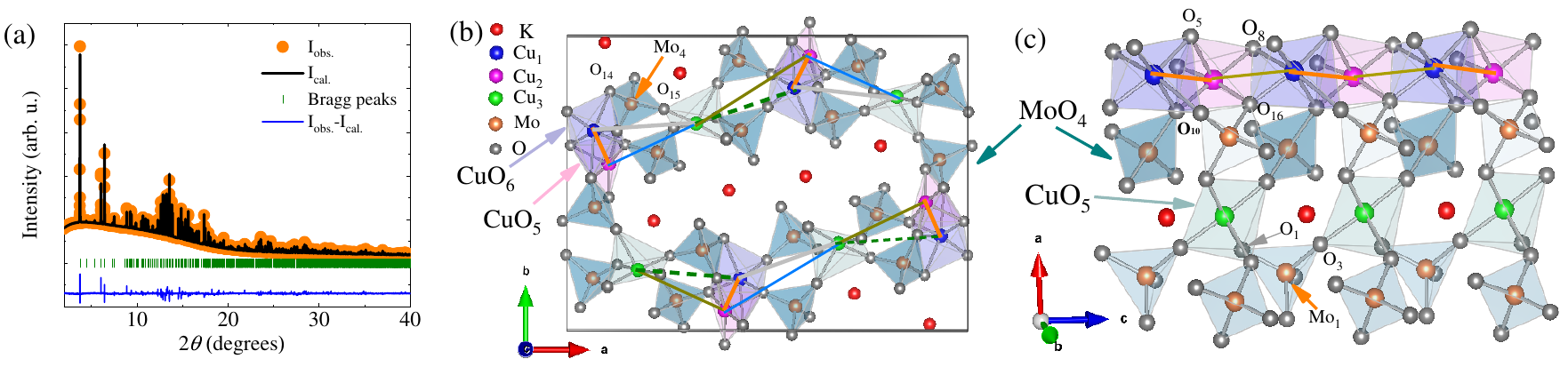}
	\caption{ (a) Rietveld refinement of synchrotron XRD powder pattern of K$_2$Cu$_3$(MoO$_4$)$_4$  measured at room temperature.  The observed data (I$_{\rm obs.}$), Rietveld refinement fit (I$_{\rm cal.}$), Bragg peaks, and difference curve (I$_{\rm obs.}$ $-$ I$_{\rm cal.}$) are denoted by the filled circle symbols, black solid line, vertical olive bars, and blue line, respectively. (b) One unit cell  of monoclinic crystal structure of K$_2$Cu$_3$(MoO$_4$)$_4$ viewed perpendicular to the $c$-axis. Alternating dimers (Cu$_{2}$O$_{9}$) and monomers of Cu$^{2+}$ ions share corners with tetrahedral MoO$_4$ units, forming a zigzag distorted diamond chain along the $a$-direction. (c) Schematic shows a chain of alternating CuO$_{6}$ octahedra (Cu${1}$ site) and CuO$_{5}$ distorted square pyramids  (Cu${2}$ site) along the $c$-axis, with isolated CuO$_{5}$ distorted square pyramids  located at the Cu${3}$ sites.  }{\label{st}}.
\end{figure*}
\section{RESULTS AND DISCUSSION}
\subsection{Rietveld refinement and crystal structure }
In order to confirm phase purity and crystal structure of polycrystalline samples of K$_2$Cu$_3$(MoO$_4$)$_4$,  Rietveld refinements of SXRD data were performed using FullProf suite software \cite{rodriguez1990fullprof}. 
Figure~\ref{st}(a) shows the Rietveld refinement of SXRD powder pattern of K$_2$Cu$_3$(MoO$_4$)$_4$, confirming the absence of any detectable secondary phase in the investigated compound. The refinements suggest the titled compound crystallizes in a monoclinic structure with space group of $P2_{1}/a$ (No.14). The  obtained lattice parameters and goodness of fit are tabulated in table~\ref{table}, which are in good agreement with previous report (ICSD \#20767)  \cite{sm}.\\
 \begin{table}[b]
	\caption{\label{table} Crystallographic parameters of K$_{2}$Cu$_{3}$(MoO$_{4}$)$_{4}$ from a  Rietveld refinement of synchrotron x-ray diffraction data at room temperature. ($\chi^{2}$ = 4.57, Space group: $P 2_{1}/a,$ $a$ = 20.072 \AA , $b$ = 14.673 \AA, $c$ = 5.144 \AA, $ $ $\alpha$ = $\gamma = 90^{\circ}$, $ \beta$ = 94.570$^{\circ}$.)} 
		\begin{tabular}{c c c c c  c c} 
		\hline \hline
		Atom & Wyckoff position & \textit{x} & \textit{y} &\textit{ z}& Occ.\\
		\hline 
		Mo$_{1}$ & 4$e$ & -0.065 \ \ & 0.821 \ \ & 0.805\ \ & 1 \\
		Mo$_{2}$ & 4$e$ & -0.070 \ \ & 0.581 \ \ & 0.307\ \ & 1 \\
		Mo$_{3}$ & 4$e$ & 0.226 \ \ & 0.602 \ \ & 0.715\ \ & 1 \\
		Mo$_{4}$ & 4$e$ & 0.161 \ \ & 0.771 \ \ & 0.220\ \ & 1 \\
		Cu$_{1}$ & 4$e$ & 0.067 \ \ & 0.676 \ \ & 0.689\ \ & 1 \\
		Cu$_{2}$ & 4$e$ & 0.105	 \ \ & 0.565\ \ & 0.206\ \ & 1 \\
		Cu$_{3}$ & 4$e$ & 0.322 \ \ & 0.704\ \ & 0.243\ \ & 1 \\
		K$_{1}$ & 4$e$ & 0.217 & 0.373 & 0.253 & 1\\
		K$_{2}$ & 4$e$ & 0.404 & 0.476 & 0.308 & 1\\
		O$_{1}$ & 4$e$ & -0.101 & 0.851 & 0.096 & 1\\
		O$_{2}$ & 4$e$&  0.001 & 0.741 & 0.855 & 1 \\
		O$_{3}$ & 4$e$& -0.126 & 0.7762 & 0.576 & 1\\
		O$_{4}$ &  4$e$& -0.033 & 0.9213 & 0.689 & 1\\
		O$_{5}$ & 4$e$& 0.010 & 0.628 & 0.391& 1\\	
		O$_{6}$ & 4$e$& -0.105 & 0.547& 0.599& 1\\	
		O$_{7}$ & 4$e$& -0.123 & 0.658& 0.139& 1\\	
		O$_{8}$ & 4$e$& -0.061 & 0.483& 0.112& 1\\		
		O$_{9}$ & 4$e$& 0.208 & 0.514& 0.915& 1\\	
		O$_{10}$ & 4$e$& 0.145 & 0.631& 0.514 & 1\\
		O$_{11}$ & 4$e$& 0.266 & 0.688& 0.920 & 1\\	
		O$_{12}$ & 4$e$& 0.281 & 0.559& 0.502 & 1\\	
		O$_{13}$ & 4$e$& 0.169 & 0.859& 0.004 & 1\\	
		O$_{14}$ & 4$e$& 0.108 & 0.811& 0.440 & 1\\	
		O$_{15}$ & 4$e$& 0.242 & 0.752& 0.387 & 1\\	
		O$_{16}$ &4$e$& 0.125 & 0.676& 0.0133 & 1\\				
		\hline
	\end{tabular}
\end{table} 
\begin{table}[h]
	\centering
	\caption{\label{tableangle}Bond lengths and angles associated to a few
		superexchange interactions between Cu$^{2+}$ moments in K$_2$Cu$_3$(MoO$_4$)$_4$. }
	\begin{tabular}{c|c}
		\toprule
		Bond length (\AA) & Bond angle ($^\circ$) \\
		\midrule
		Cu${1}$-Cu${2}$ = 3.11 (Fig.\ref{st}(c)) & Cu${2}$-O${5}$-Cu${1}$ = 90.36 \\
		& Cu${2}$-O${10}$-Cu${1}$ = 103.94 \\ \hline
			Cu${2}$-Cu${1}$ = 3.15 (Fig.\ref{st}(c)) & Cu${2}$-O${8}$-Cu${1}$ = 87.72 \\
		& Cu${2}$-O${16}$-Cu${1}$ = 107.5 \\ \hline
		
			Cu${3}$-Cu${3}$ = 5.14 (Fig.\ref{st}(c)) & Cu${3}$-O${1}$-Mo${1}$ = 128.2 \\
		& Mo${1}$-O${3}$-Cu${3}$ = 145.3 \\ \hline
		
		Cu${1}$-Cu${3}$ = 5.78 (Fig.\ref{st}(b)) & Cu${1}$-O${14}$-Mo${4}$ = 107.8 \\
		& Mo${4}$-O${15}$-Cu${3}$ = 127.9 \\ \hline
		
			Cu${2}$-Cu${3}$ = 4.79 (Fig.\ref{st}(b)) & Cu${2}$-O${16}$-Mo${4}$ = 114.6 \\
		& Mo${4}$-O${15}$-Cu${3}$ = 127.9 \\ \hline
		\bottomrule
	\end{tabular}
\end{table}
One unit cell of monoclinic structure of K$_2$Cu$_3$(MoO$_4$)$_4$ is depicted in Fig.~\ref{st}(b). In this structure, copper atoms are distributed across three distinct crystallographic sites: Cu1 (CuO$_{6}$ octahedra), Cu2, and Cu3 (distorted square pyramids). Notably, Cu1 and Cu2 atoms share an edge, forming the dimer unit Cu$_2$O$_9$, while Cu3 exists as a monomeric distorted square CuO$_5$ unit (see Fig.~\ref{st}(b)). In the $ab$-plane, these copper atoms, arranged alternately as monomers and dimers, are linked by corner-shared tetrahedral MoO$_4$ units, creating an infinite zigzag distorted diamond chain along the $a$-direction. Furthermore, the adjacent chains are connected by tetrahedral MoO$_4$ units to establish an exchange path for  inter-chain interactions, with K$^+$ ions occupying interstitial sites between the chains. The dashed red line in Fig.~\ref{st}(b)  depicts the arrangement of Cu$^{2+}$ ions, forming a network of distorted diamond chain.\\From a structural point of view, the titled compound K$_2$Cu$_3$(MoO$_4$)$_4$ shares a  similar space group symmetry with the previously reported diamond chain compound Cu$_{3}$(CO$_{3}$)$_{2}$(OH)$_{2}$ \cite{PhysRevLett.94.227201}.
	However, in Cu$_{3}$(CO$_{3}$)$_{2}$(OH)$_{2}$ (azurite), there are two crystallographically distinct Cu$^{2+}$ sites, labeled Cu1 and Cu2. The Cu1 sites form dimers that repeat along the $b$-axis with the same orientation, while the monomeric Cu2 spins are interspersed between them. As a result, the system is characterized by three exchange interactions: $J_{1}$, $J_{2}$, and $J_{3}$ \cite{PhysRevLett.94.227201,PhysRevB.83.104401}.
In contrast, K$_2$Cu$_3$(MoO$_4$)$_4$ features three distinct Cu$^{2+}$ sites. Here, Cu1 and Cu2 also form dimers, but unlike azurite, these dimers repeat along the $a$-axis with alternating orientations (see Fig.~ \ref{st}(b)). Additionally, the distances between the monomeric Cu3 sites and the dimers vary. As a result, this system forms a distorted dimer chain, necessitating at least five distinct exchange interactions for a minimal description of magnetism, as discussed in Sec. \ref{DFT}.\\
Most notably, the aforementioned diamond magnetic structure along the crystallographic 
$a$-axis in K$_2$Cu$_3$(MoO$_4$)$_4$ is expected to dictate a lower-energy magnetic behavior, as the interactions between dimers and monomers are mediated through the superexchange path Cu--O--Mo--O--Cu (Table~\ref{tableangle} and Fig.~\ref{st}(c)). Conversely, higher-energy spin dynamics are expected along the crystallographic 
$c$-axis, as the dimers  (dashed line in Fig.~\ref{st}(c)) are via the Cu--O--Cu superexchange path and form a 1D chain. This dimer chain is connected to the chain of monomers through Cu--O--Mo--O--Cu, providing inter-chain interactions in the $ac$-plane (Fig.~\ref{st}(c)). The differing energy scales along the different axes may lead to dichotomous magnetic behaviors in the two magnetic subsystems of K$_2$Cu$_3$(MoO$_4$)$_4$ similar to that observed in the $s$ = 1/2 diamond chain K$_{3}$Cu$_{3}$AlO$_{2}$(SO$_{4}$)$_{4}$ \cite{doi:10.7566/JPSJ.84.073702}. Before proceeding, we admit that while the diamond spin chain appears evident from the crystal structure, real candidate materials involving intriguing super-super exchange paths often exhibit complexities in their spin networks. In Sec.\ref{DFT}, we will address this issue through DFT calculations    
\subsection{Magnetic susceptibility}
 \begin{figure*}
	\centering
	\includegraphics[width=\textwidth]{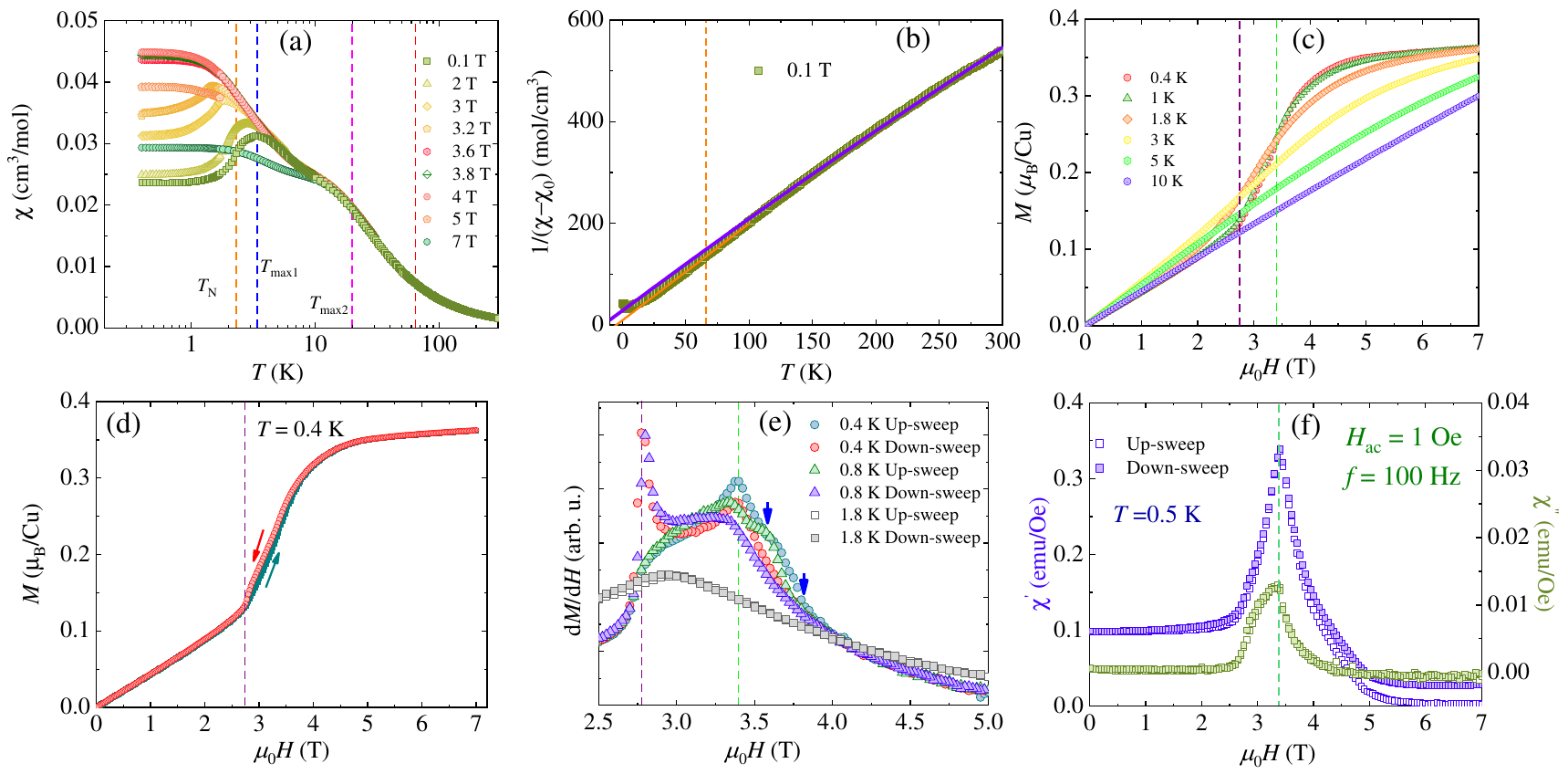}
	\caption{(a) Temperature dependence of magnetic susceptibility ($\chi(T)$) for K$_2$Cu$_3$(MoO$_4$)$_4$  in a semi-logarithmic scale under several magnetic  fields. The orange, blue, pink, and red dashed vertical lines represent the positions of transition, two broad maxima, and crossover temperatures, respectively,  at $\mu_{0}H$ = 0.1 T, as described in the main text. (b) Temperature dependence of inverse magnetic susceptibility (1/$\chi$) after subtracting the diamagnetic contributions,  where the violet, and orange solid lines represent the Curie-Weiss fits in the two distinct regimes. 
		(c) Isothermal magnetization as a function of magnetic field at various temperatures for only up-sweep of fields, and in (d) both up-sweep and down-sweep of fields at a 0.4 K. (e) Derivative of magnetization for both up-sweep and down-sweep of fields at three different temperatures. The beginning and end of additional features observed exclusively during the up-sweep of fields are indicated by a blue arrow at 0.5 K. (f) Magnetic field dependence of the real part (left $y$-axis) and imaginary part (right $y$-axis) of \textit{ac} susceptibility for both up-sweep and down-sweep of fields at 0.5~K.	The dashed vertical lines in Fig.~(c-f)	indicate a metamagnetic transition at 2.6 T (purple line) and a critical field at 3.4 T (green line), \textcolor{blue}{above} which the AFM order becomes suppressed.}{\label{chi}}.
\end{figure*}
\begin{figure*}
	\centering
	\includegraphics[width=\textwidth]{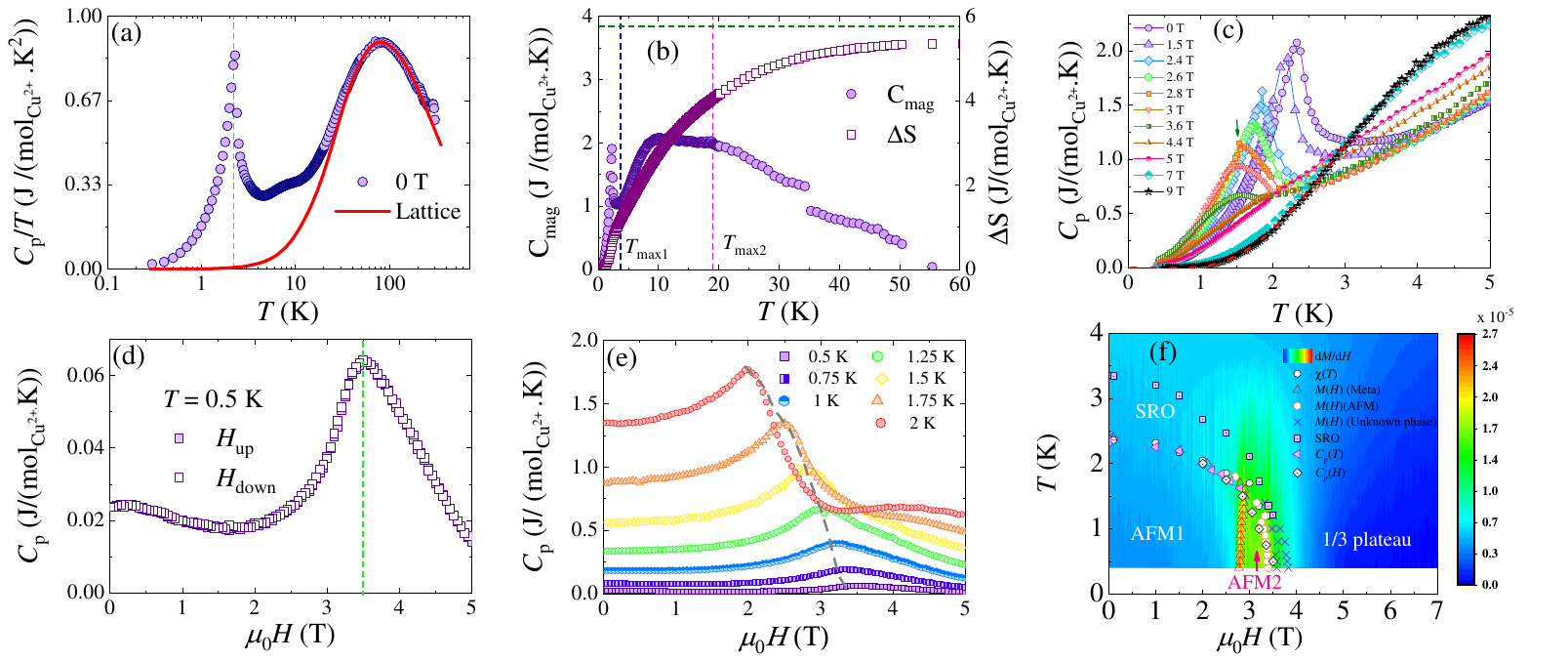}
	\caption{(a) Temperature dependence of specific heat ($C_{p}$) divided by temperature for K$_2$Cu$_3$(MoO$_4$)$_4$ in zero field. The solid red line represents the lattice contribution obtained by combining one Debye and three Einstein functions as described in the main text. (b) Temperature dependence of magnetic specific heat ($C_{\rm mag}$; left $y$ axis) and the calculated entropy  change ($\Delta S$;, right $y$ axis).  (c) Low temperature  $C_{p} $ as a function of temperature in several magnetic fields in the temperature range 0.3 K $\leq$ $T$ $\leq$ 5 K. (d) $C_p$ as a function of the magnetic field for both up- and down-sweeps at $T = 0.5$~K, and (e) for the up-sweep only at various temperatures. The dashed dark yellow line guides the eye to the shift of the $\mu_{0}H_{c}$. (f) Temperature versus magnetic field phase diagram where the phase boundaries were determined using various methods, as mentioned in the legends. In the phase diagram, SRO stands for short-range order, while AFM1 and AFM2  correspond to the antiferromagnetic phases. The dashed vertical lines have the same interpretation as in Fig.~\ref{chi}.    }{\label{HC}}.
\end{figure*}

 
Figure~\ref{chi}(a) shows the temperature dependence of magnetic susceptibility for  K$_2$Cu$_3$(MoO$_4$)$_4$ in a semi-logarithmic scale in several magnetic fields.  As the temperature decreases, $\chi(T)$ gradually increases. However, below 65 K, where high-temperature Curie-Weiss fit deviates from the measured susceptibility data (red dashed line  in Figs.~\ref{chi}(a) and \ref{chi}(b)), it rapidly increases, followed by a broad maximum around 20 K ($T_{\rm max2}$ in Fig.~\ref{chi}(a)), corresponding to the characteristic feature of a spin-singlet state associated with the Cu$_{2}$O$_{9}$ dimer. Upon further lowering temperature, $\chi(T)$ exhibits another pronounced broad maximum around 3.68 K at $\mu_{0}H$ = 0.1 T (${T_{\rm max1}}$ in Fig.~\ref{chi}(a)), alluding to the presence of short-range spin correlations due to the effective magnetic interactions between structural dimer and monomers in  K$_2$Cu$_3$(MoO$_4$)$_4$. Below $T_{\rm max1}$, the decrease of $\chi(T)$ indicates the development of AFM spin correlations, which eventually lead to a long-range magnetic order at $T_{\rm N}$ = 2.3 K as evidenced by specific heat data (see section \ref{spec}).   The absence of any Curie-like upturn below $T_{\rm N}$ suggests the high purity of the polycrystalline sample used in this study \cite{doi:10.7566/JPSJ.84.073702}. 
Similar to K$_2$Cu$_3$(MoO$_4$)$_4$, the diamond chain compound Cu$_3$(CO$_3$)$_2$(OH)$_2$ exhibits a double-peak feature with maxima at 20 K and 5 K above its Néel temperature of $T_{\rm N}$ = 1.8 K \cite{PhysRevLett.94.227201}. \\
 As the magnetic field strength increases, the position of \( T_{\mathrm{max2}} \) remains unchanged up to the highest field \( \mu_{0}H = 7 \ \mathrm{T} \). In contrast, \( T_{\mathrm{max1}} \) remains stable up to \( \mu_{0}H = 2 \ \mathrm{T} \), but begins to shift to lower temperatures as the field exceeds \( \mu_{0}H = 2 \ \mathrm{T} \). Notably, at \( \mu_{0}H = 3 \ \mathrm{T} \), \( \chi(T) \) continues its decreasing trend below $T_{\rm max1}$, indicating that AFM interactions still dominate at this field strength. Interestingly, at \( \mu_{0}H = 4 \ \mathrm{T} \), \( \chi(T) \) displays a step-like increase immediately after \( T_{\mathrm{max2}} \), instead of following the usual behavior, and then saturates at low temperatures. This behavior suggests the field-induced suppression of AFM order, giving way to a partially polarized state. Furthermore, as the magnetic field is increased beyond \( \mu_{0}H \geq 4 \ \mathrm{T} \), \( \chi(T) \) saturates at a lower values below \( T_{\mathrm{max2}} \). This field range from 4 T to 7 T coincides with the onset of the 1/3 magnetization plateau discussed below.
 \\  To obtain a rough estimate of the dominant magnetic exchange interactions between Cu$^{2+}$ moments, the  inverse magnetic susceptibility (1/$\chi(T)$) data were fitted using the Curie-Weiss law,  i.e., $\chi$ = $C/(T-\theta_{\rm CW})$, in the high-temperature linear regime. Here, $\theta_{\rm CW}$ is the Curie-Weiss temperature that offers an average magnetic exchange interaction between magnetic ions, and $C$ is the Curie constant which determines  the effective magnetic
moment ($\mu_{\rm eff}$ = $\sqrt{8C}$). It is important to note that before fitting the Curie-Weiss law, the total diamagnetic susceptibility from all constituent ions in K$_2$Cu$_3$(MoO$_4$)$_4$,  $\chi_{0} = -2.82 \times 10^{-4}$ cm$^{3}$/mol, was subtracted \cite{Bain2008}.
 As shown in Fig.~\ref{chi}(b), the 1/$\chi(T)$ data exhibits two Curie-Weiss regimes with  $\theta_{\rm CW} = - 25 $ K and $\mu_{\rm eff}$ = 1.91 $\mu_{\rm B}$ (purple line;  $C$ = 0.46 cm$^{3}$ K/mol, 100 K $\leq$ $T$ $\leq$ 300 K), and $\theta_{\rm CW} = -3.82$ K and $\mu_{\rm eff}$ = 1.87 $\mu_{\rm B}$ (orange line; $C$ = 0.44 cm$^{3}$ K/mol 40 K $\leq$ $T$ $\leq$ 100 K). Above 100 K, the negative value of $\theta_{\rm CW}$ suggests the presence of dominant antiferromagnetic magnetic interaction between Cu$^{2+}$ moments \cite{PhysRevB.91.014423}, which is attributed to the presence of a low-dimensional spin topology with antiferromagnetic interactions. 
 	 However, upon  lowering the sample temperature, the development of additional weak interactions, likely between dimers and monomers as well as three-dimensional interactions, becomes evident through the reduced negative value of  $\theta_{\rm CW}$. This reduction of the Curie-Weiss temperature at lower temperatures suggests the presence of at least ferromagnetic interactions in addition to the dominant antiferromagnetic interactions  in K$_2$Cu$_3$(MoO$_4$)$_4$.   Such a coexistence of ferromagnetic and AFM interactions has been theoretically proposed in distorted diamond compounds \cite{10.1063/9.0000255} and  observed in another distorted diamond compound, K$_{3}$Cu$_{3}$AlO$_{2}$(SO$_{4}$)$_{4}$ \cite{doi:10.7566/JPSJ.84.073702}.\\ At elevated temperatures, weak interactions remain thermally quenched, but as the temperature decreases, these interactions become increasingly significant. Typically, such systems exhibit low-dimensional behavior down to a characteristic temperature, below which weak interchain couplings or interactions between dimers and monomers become active, leading to a modification of the spin network and altering spin correlations \cite{Subir,Lake2005}. \\     
To further investigate the field-induced magnetic properties of K$_2$Cu$_3$(MoO$_4$)$_4$, isothermal magnetization measurements were conducted at various temperatures, as shown in Fig.~\ref{chi}(c). The linear behavior of \( M(H) \) up to a magnetic field of \( \mu_{0}H = 2.6 \ \mathrm{T} \) (indicated by the purple dashed line in Fig.~\ref{chi}(d)) suggests the absence of a spin gap in the zero-field ground state and the dominance of AFM interactions, consistent with the behavior of \( \chi(T) \) up to \( \mu_{0}H = 2.6 \ \mathrm{T} \). However, as the magnetic field increases beyond \( \mu_{0}H \geq 2.6 \ \mathrm{T} \), the magnetization rises rapidly at temperatures below $T$ $<$ \( T_{\rm N} = 2.3 \ \mathrm{K} \) and continues up to 4 T. For fields beyond 4 T, the magnetization gradually saturates, approaching a value close to 
0.35 $\mu_{\rm B}$, as the field increases to 7 T (see Fig.~\ref{chi}(d)). This finding indicates the presence of a plateau at approximately 1/3 of the saturation magnetization, similar to that observed in the 1D diamond chain compound Cu$_3$(CO$_3$)$_2$(OH)$_2$ \cite{PhysRevLett.94.227201}. The weak temperature dependence of the 1/3 plateau is attributed to the presence of anisotropic magnetic interactions \cite{10.1063/9.0000255}.   However, future magnetization measurements at higher fields are necessary to investigate the influence of anisotropic interactions on the width of the 1/3 magnetization plateau and the occurrence of a 5/6 plateau, which arises from the breaking of spin dimers. One possible scenario of the emergent 1/3 plateau is based on the decoupling of weakly coupled monomer spins from the spin dimers, resulting in the polarization of the monomer spins along an applied field direction. Alternatively, our DFT calculations (see section \ref{DFT}) reveal that the Cu$^{2+}$
ions form  an anisotropic triangular lattice with competing ferromagnetic and AFM interactions, which could lead to fragmentation of the origin spin lattice into subsystems under a magnetic field.  \\ 
To gain deeper insight into the rapid increase in magnetization in the field range \( 2.6 \ \mathrm{T} \leq \mu_{0}H \leq 4 \ \mathrm{T} \), the derivative of the magnetization data for both the up-sweep and down-sweep fields is shown in Fig.~\ref{chi}(e). This derivative clearly reveals a hysteresis within the field range \( 2.6 \ \mathrm{T} \leq \mu_{0}H \leq 4 \ \mathrm{T} \), indicating the occurrence of a metamagnetic transition \cite{doi:10.7566/JPSJ.86.074706,LOVE1970290}. 
This behavior indicates that the strong magnetic field overcomes the relatively weak spin anisotropy, causing the antiferromagnetically ordered spins to reorient perpendicular to the field \cite{white1983quantum}. Additionally, the \( dM/dH \) data shows two anomalies, one at 2.6 T and another at 3.4 T. The anomaly at 2.6 T is attributed to the field-induced metamagnetic transition to  a field-induced antiferromagnetic phase (AFM2), while the anomaly at 3.4 T  corresponds to the critical magnetic field ($\mu_{0}H_{c}$) above which the AFM order is suppressed due to the decoupled spin dimers from the polarized spin monomers \cite{Bera2020,Wu2019}. Most importantly, between the $\mu_{0}H_{c}$ = 3.4 T and just before the 1/3 plateau state, an additional hump appears (blue arrow in Fig.~\ref{chi} (e)) only during the up-sweep of the magnetic field and persists up to 1.5 K, suggesting the presence of an additional phase  between the $\mu_{0}H_{c}$ and the 1/3 plateau phase.  However, the nature of this additional phase remains unclear in this study. In the phase diagram shown in Fig.~\ref{HC}(f), we label this additional phase as an ``unknown phase."  In low-dimensional magnets, an  additional phase between $\mu_{0}H_{c}$ and the 1/3 magnetization plateau has been identified as an incommensurate longitudinal spin-density wave (LSDW) state, as observed in YbAlO$_{3}$ \cite{Wu2019,mokht}.  \\
To determine the critical field more precisely, ac susceptibility measurements were conducted at 0.5~K, as shown in Fig.~\ref{chi}(f). A clear peak in both the real ($\chi'$) and imaginary ($\chi''$) components of the ac susceptibility at 3.4~T provides further evidence for the presence of a $\mu_{0}H_{c}$ around this field. No additional features were observed in the ac susceptibility near the metamagnetic transition at 2.6~T; however, both $\chi'$ and $\chi''$ exhibit a sharp increase just below the metamagnetic transition at 2.6~T. The bifurcation in $\chi'$ between the up-sweep and down-sweep of the magnetic field is likely associated with  an additional phase, which also appears in the field range  3.5~T to 3.8~T in $dM$/$dH$ (Fig.~\ref{chi}(e)).\\
It is noteworthy that field dependence of $\chi^{'}$ and $\chi^{''}$ (Fig.~\ref{chi}(f)) do not exhibit hysteresis at 3.4 T, whereas  $dM/dH$ (Fig.~\ref{chi}(e)) does. One possible explanation is that ac susceptibility measurements probe the system’s response to an alternating magnetic field and are typically more sensitive to dynamical properties, such as relaxation and dissipation. At the critical field of 3.4 T, where the system undergoes a transition between states, the alternating field may not induce a sufficiently strong dynamic response to reveal hysteresis in $\chi^{'}$ and $\chi^{''}$. ac susceptibility measurements can smooth out the effects of slower transitions or irreversible processes, which may account for the absence of hysteresis in this case.
\subsection{\label{spec}Specific heat}
To gain more insight into the magnetic properties, we performed specific heat ($C_{p}$) measurements of K$_2$Cu$_3$(MoO$_4$)$_4$ in the temperature range 0.3 K $\leq$ $T$ $\leq$ 300 K in several magnetic fields. Figure~\ref{HC}(a) shows the temperature dependence of $C_p$/$T$  in  zero field. A sharp lambda-type anomaly at $T_{\rm N}$ =  2.3 K confirms the presence of a long-range magnetic  order state.\\ In magnetic insulators, the measured $C_p$ has two principal contributions: the phononic and magnetic part that dominate at higher and lower temperatures, respectively. To estimate the phonon contributions, we fitted the $C_p$ data in the temperature range 60 K $\leq$ $T$ $\leq$ 150 K (red line in Fig.~\ref{HC}(a)) using the formula consisting of a linear combination of one Debye and three Einstein terms \cite{kittel2018introduction,Khatua2021} as
\begin{equation}
\begin{split}
C_p^{ph}(T) = C_{\rm D} \left[ 9R \left(\frac{T}{\theta_{\rm D}}\right)^3 
\int_0^{\theta_{\rm D}/T} \frac{x^4 e^x}{(e^x-1)^2} \, dx \right] \\
+ \sum_{i=1}^{3} C_{\rm E_{i}} \left[ R \left(\frac{\theta_{\rm E_{i}}}{T}\right)^2 
\frac{\exp\left(\frac{\theta_{\rm E_{i}}}{T}\right)}{\left[\exp\left(\frac{\theta_{\rm E_{i}}}{T}\right)-1\right]^2} \right],
\end{split}
\end{equation}
where, $\theta_{\rm D}$ is the Debye temperature, $\theta_{\rm E_is}$ are the Einstein temperatures of the three modes, and $R$ represents the molar gas constant. The fitting parameters are 
$\theta_{\rm D}$ = 74(1) K, $\theta_{\rm E_1}$ = 146(3) K, $\theta_{\rm E_2}$ = 304(2) K, and $\theta_{\rm E_3}$ = 633(6) K. To minimize the fitting parameters, the coefficients were set in the fixed ratio of 
\( C_{\rm D} : C_{\rm E_1} : C_{\rm E_2} : C_{\rm E_3} = 3 : 2 : 8 : 12 \), 
corresponding to the ratio of three acoustic and the \((3n - 3)\) optical phonon branches, where \( n \) is the total number of atoms in K\(_{2}\)Cu\(_{3}\)(MoO\(_{4}\))\(_{4}\) \cite{PhysRevB.63.212408,PhysRevB.110.184402}.
\\
After subtracting the lattice contribution, the obtained temperature dependence of the magnetic specific heat, $C_{\rm mag}$, is depicted in Fig.~\ref{HC}(b). It is observed that below 60 K, $C_{\rm mag}$ starts to increase and shows a broad maximum around 20 K, consistent with the $\chi{(T)}$ data \cite{PhysRevB.103.064413,Khatua2022,PhysRevMaterials.5.054413}.  Furthermore, the lack of a clear distinction between the broad maxima from spin-singlet and short-range correlations is attributed to the overlap of both contributions in $C_{\rm mag}$.  Below 10 K, $C_{\rm mag}$ begins to decrease until  2 K, which may be the expected limit of  spin-singlet correlations. However, below 2 K, $C_{\rm mag}$ starts to increase again, indicating the development of three-dimensional AFM spin correlations. This is evidenced by a lambda-type anomaly at $T_{\rm N}$ = 2.3 K, suggesting the onset of a long-range ordered state.\\ To estimate the percentage of entropy released at $T_{N}$ and above, the entropy was calculated by integrating $C_{\rm mag}/T$ with respect to temperature, as depicted on the right $y$-axis of Fig.~\ref{HC}(b). The estimated $\Delta$S$_{\rm mag}$ saturates at 55 K to a value of 5.35 J/mol$\cdot$K, consistent with the theoretical value $R$ ln(2$s$ $+$ 1) = 5.76 J/mol$\cdot$K for $s$ = 1/2. The spin entropy recovered at $T_N$ is approximately 0.64 J/mol$\cdot$K, corresponding to roughly 12\% of the total spin entropy. These results indicate that nearly 88\% of the spin entropy is released above $T_N$ due to short-range spin correlations.\\
To understand the effect of an applied magnetic field on the magnetic ground state, specific heat measurements were performed down to 0.3 K in several magnetic fields  as shown in Fig.~\ref{HC}(c) as a function of temperature. Upon increasing magnetic field, the lambda-type peak shifts to lower temperatures, which is a characteristic behavior observed in AFM materials undergoing a magnetic transition. It is observed that the lambda-type peaks persist up to at least $\mu_{0}H = 2.4$ T; however, at 2.6 T, a broad peak emerges at lower temperatures, coinciding with the metamagnetic transition observed in the magnetization data. Therefore, such broad feature above 2.4 T suggests an overlap of anomalies associated with the metamagnetic and AFM transitions. Upon further increasing magnetic field, the $C_p$ data broadens further, making it challenging to distinguish specific features due to the metamagnetic transition. However, at 2.8 T, a weak kink can only be seen, marked by an olive arrow, as the indication of metamagnetic transition (see Fig.~\ref{HC} (c)). It is noteworthy that a weak broad feature persists even at 3.6 T, beyond the $\mu_{0}H_{c}$. This broad anomaly aligns with the feature observed in the $dM/dH$ data in the field range of 3.5 T to 3.8~T (Fig.~\ref{chi}(e)).  
 Interestingly, for fields $\mu_{0}H > 3.6$~T, these weak broad features are suppressed, and the $C_p$ data begin to exhibit exponential behavior at low temperatures, indicating that the singlet-triplet gap from Cu$_{2}$O$_{9}$ dimers is gradually becoming distinguishable. As the magnetic field increases, the exponential behavior becomes more pronounced at 9~T (dashed red line in Fig.~\ref{HC}(c)), suggesting the decoupling of monomer spins from the Cu$_{2}$O$_{9}$ dimers.\\
 \begin{figure}
 	\centering
 	\includegraphics[width=0.46\textwidth]{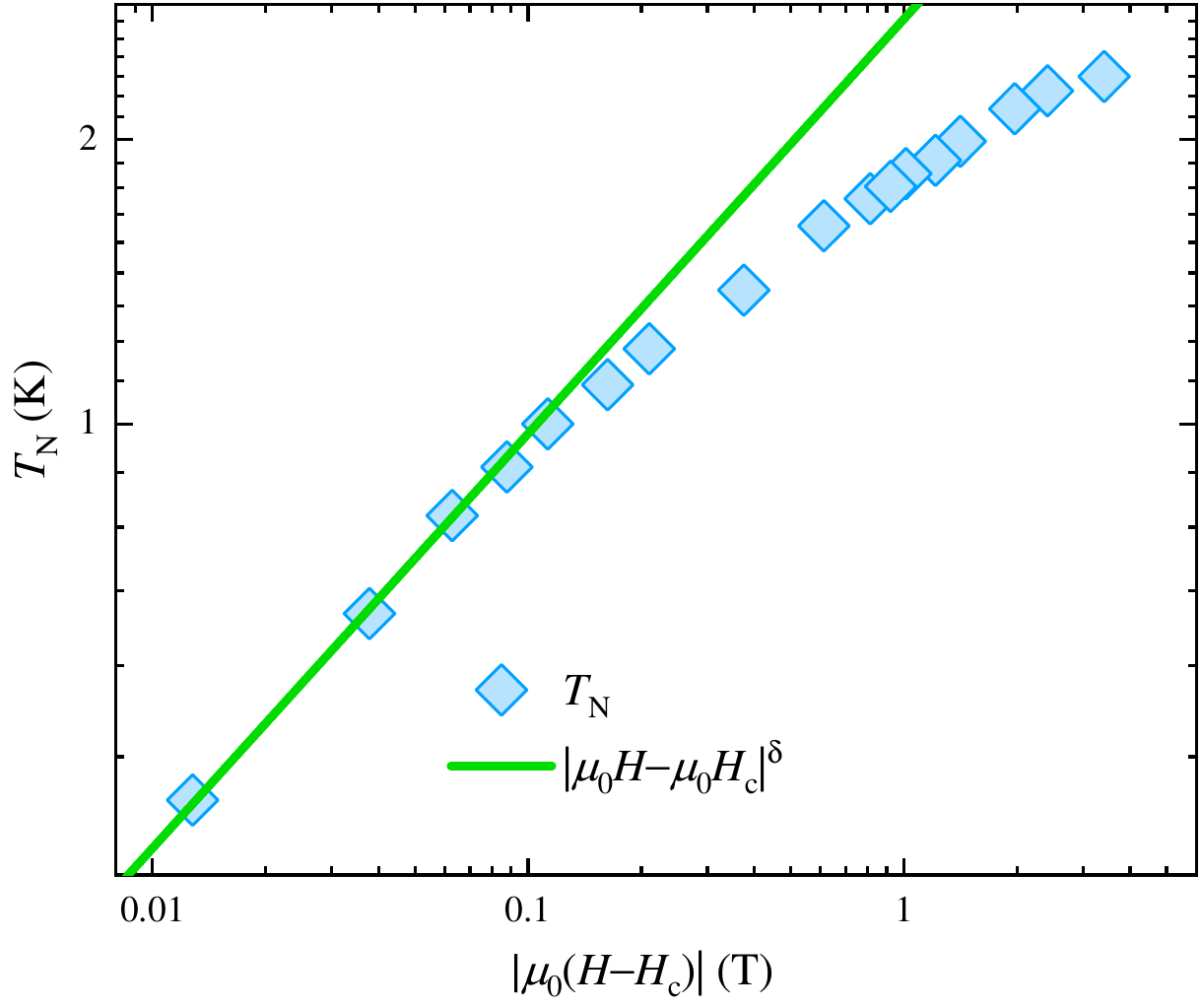}
 	\caption{$T_{\rm N}$  as a function of $|\mu_{0}H-\mu_{0}H_{c}|^{\delta}$, with $\mu_{0}H_{c}$ = 3.4 T on a logarithmic scale. The solid green line represents $T_{N}\propto $ $|\mu_{0}H-\mu_{0}H_{c}|^{0.46}$.   }{\label{delta}}.
 \end{figure}
Fitting the $C_p$ data at 9~T to an exponential form $\exp(\Delta / k_B T)$ yields an intra-dimer  spin gap of $\Delta/k_{\rm B} = 6.97$~K.  As this gap is a rough estimation, future high-field magnetization measurements could be helpful in extracting the exact intra-dimer gap.  \\  To further investigate the field-induced $\mu_{0}H_{c}$, specific heat was measured as a function of magnetic field at various temperatures, as shown in Figs.~\ref{HC}(d) and \ref{HC}(e). Unlike the magnetization data, no bifurcation between the up-and down-sweep of fields was observed, indicating a weak hysteresis in the magnetization (Fig.~\ref{HC}(d)). Instead, a peak around 3.4~T suggests the presence of a $\mu_{0}H_{c}$, consistent with the magnetization data (Fig.~\ref{chi}). As the temperature increases, the peaks shift to lower fields (dashed line in Fig.~\ref{HC}(e)), in agreement with the magnetization results.\\ 
Taking into account the field dependence of
anomalies in magnetic susceptibility, isothermal magnetization, and specific heat data, we constructed a temperature-field phase diagram, as shown in Fig.~\ref{HC}(f). This diagram clearly illustrates the suppression of AFM order with increasing applied magnetic field strength. At a critical field of approximately 3.4 T, the AFM order is completely suppressed. Below the $\mu_{0}H_{c}$, the field-induced antiferromagnetic phase is denoted as AFM2, which is separated from the zero field antiferromagnetic phase (AFM1) by a metamagnetic transition at 2.6 T.  It also reveals the presence of an additional phase labeled as an``unknwon phase" just before 1/3 plateau phase. To characterize the distinct phases on either side of the $\mu_{0}H_{c}$, future neutron diffraction studies are essential.\\In order to understand the nature of the phase transition near $\mu_{0}H_{c}$, we have shown  that the transition temperature follows a universal power-law behavior $T_{\rm N} \propto|\mu_{0}H-\mu_{0}H_{c}|^{\delta}$,   with $\delta$ = 0.46(2). This value is close to the expected $\delta$ = 0.66 for Bose-Einstein condensation of magnetic excitations in three-dimensional magnets near the critical field or the mean-field value of 1/2 \cite{Sebastian2006}.
\subsection{Electron spin resonance} 
\begin{figure*}
 	\centering
 	\includegraphics[width=\textwidth]{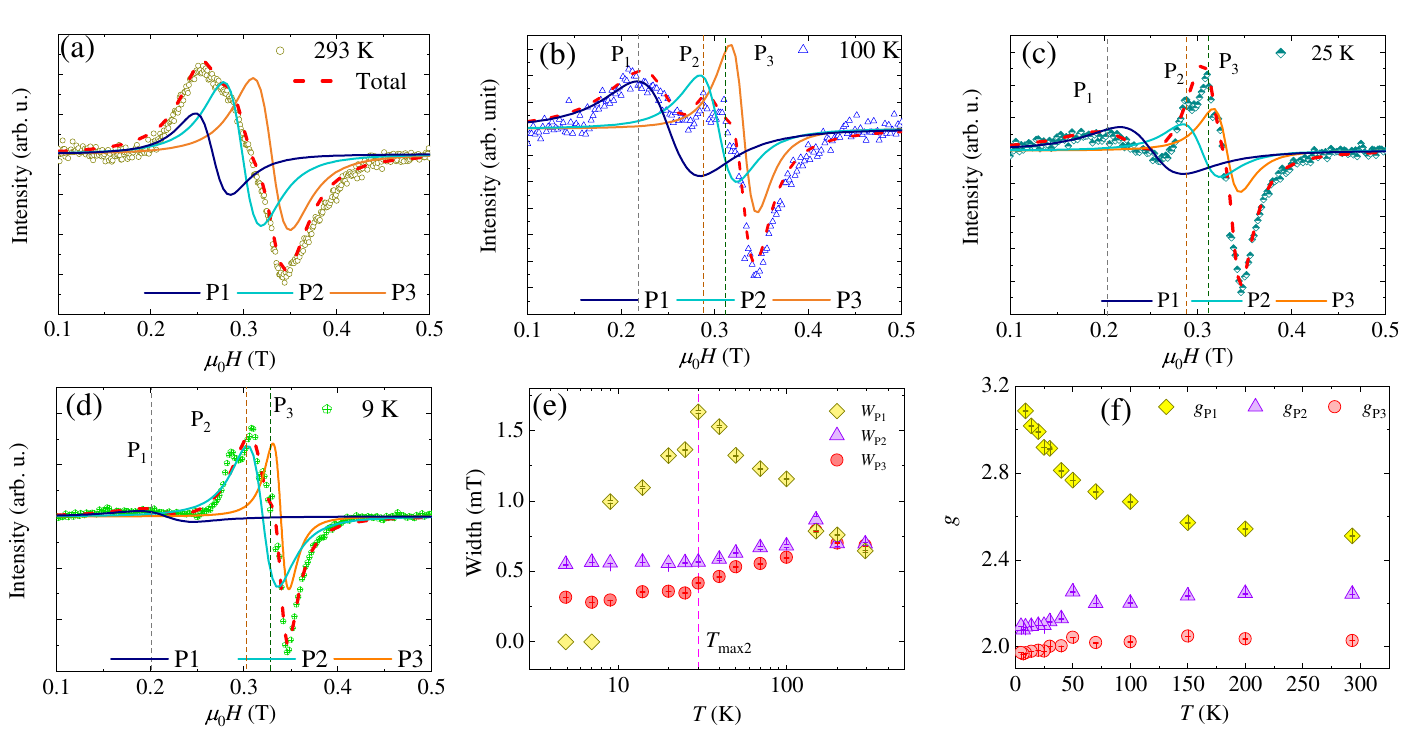}
 	\caption{(a-d) ESR spectra at several selected temperatures.  Solid lines represent each individual Lorentzian peak (P1, P2, and P3) separately, while the dashed line corresponds to the total fit. The lowest-field peaks are indicated by dashed vertical lines. (e) Temperature dependence of the linewidth for the three Cu$^{2+}$ sites. The dashed dashed vertical pink line carries the same meaning as in the $\chi(T)$ and $C_{\rm  mag}$ data. (f) Temperature dependence of the $g$-factors for each line. Error bars for linewidths and g-factors  are smaller than the symbol size. The small maximum observed in the positive derivative, which appears less pronounced in the negative derivative, is likely attributable to background contributions rather than a distinct spectral feature.  }{\label{ESR}}.
 \end{figure*}

To gain further microscopic insights into  electron spin
correlations
 in K$_{2}$Cu$_{3}$(MoO$_{4}$)$_{4}$,
 ESR measurements were performed at a constant microwave
 frequency  $\nu$ = 9.5
 GHz while varying the external static magnetic field $\mu_{0}H$. In our experimental setup, the lowest measured temperature was 5 K which is higher than the onset of long-range AFM order at $T_{\rm N}$ = 2.3 K  for K$_{2}$Cu$_{3}$(MoO$_{4}$)$_{4}$.\\  Figure~\ref{ESR}(a) to \ref{ESR}(d) present the obtained first derivative of ESR spectra at several temperatures.  The asymmetric spectrum at room temperature indicates the presence of powder-averaged  broadening due to a convolution of multiple ESR absorption lines \cite{HBenner1983}.
 The asymmetric resonance line is well described by three Lorentzian functions (red dashed line in Fig.~\ref{ESR}(a)), which may originate from different Cu$^{2+}$ crystal field environments, leading to distinct $g$-factors.
 As seen from structural characterization, there are two crystal field environments of Cu$^{2+}$ ions: (i) CuO$_{6}$ octahedra (Cu${1}$ site) and (ii) CuO$_{5}$ distorted square pyramides (Cu${2}$ and Cu${3}$ sites). The distinct environment of Cu$^{2+}$ ions results in different site symmetries, typically resulting in two distinct $g$-factors for each of the octahedral and square pyramidal crystal fields. Two of the four $g$-values may be similar, allowing for the resolution of three resonance lines in our power-averaged spectra.   \\
At temperatures $T \gg 100$ K, the spectra in the paramagnetic regime do not clearly distinguish the three resonance lines but a distinct separation of the three resonance lines becomes pronounced as the temperature decreases further. The dashed vertical lines in Fig.~\ref{ESR} (b) denote the three peaks on the lowest field side, each representing a distinct Cu$^{2+}$ environment. For convenience, we designate the peaks as P$_{1}$, P$_{2}$, and P$_{3}$.\\ As the temperature decreases, the P$_{1}$ peak starts to broaden and disappears at 9 K, while the P$_{2}$ and P$_{3}$ peaks undergo little change down to the lowest measured temperature (Fig.~\ref{ESR}(b) to \ref{ESR}(d)). This together with the upturn in the magnetic specific heat below 9 K indicates a crossover from 1D to three-dimensional spin correlations. Figure~\ref{ESR}(e) shows the temperature dependence of the linewidth for P$_{1}$, P$_{2}$, and P$_{3}$, each showing distinct characteristic features. For example, the linewidth of the P$_{2}$ and P$_{3}$ lines shows a weak decrease down to 30 K and becomes temperature-independent in  lower temperatures. In contrast, the linewidth of the P$_{1}$ line increases monotonically with decreasing temperature, followed by a broad peak around 30 K. \\ The ESR linewidth is strongly related to spin
correlations in low-dimensional transition metal based systems \cite{doi:10.1143/JPSJ.9.888,AZorko,PhysRevB.65.134410}. As the temperature decreases, the initial increase in linewidth can be attributed to the development of spin correlations. As the ESR linewidth remains largely temperature-independent, its increase indicates coupling between the Cu$_{2}$O$_{9}$  dimers \cite{PhysRev.135.A640,PhysRevB.80.012408,Choi_2014,PhysRev.135.A640,PhysRevB.72.012415}. 
 Below 30 K, the decrease in linewidth indicates that the system transits through a thermal crossover from coupled dimers to a 3D connected state, driven by additional magnetic interactions \cite{PhysRevB.95.184430}. We recall that a broad feature in both the ESR linewidth and magnetic specific heat is characteristic of quantum magnets consisting of dimers coupled to other spins or forming random dimer networks \cite{ZORKO2004E699,PhysRevB.95.184430,Choi_2014}.
 As the temperature continues to decrease below 30 K, $W_{\rm P1}$ slowly drops to its minimum value of 1 mT at 9 K below which spectral simulation using three Lorentzian functions becomes unsuccessful. Instead, two Lorentzian functions are adequate to simulate the spectra.  The suppression of the P$_{1}$ peak at low temperatures, suggesting a competing relaxation mechanism, leading to a wipe out of the ESR signal within the X-band frequency window.     \\ 
Figure~\ref{ESR}(f) presents the estimated $g$-factors (\textit{g}$_{\rm P1}$, \textit{g}$_{\rm P2}$, and \textit{g}$_{\rm P3}$) as a function of temperature for K$_{2}$Cu$_{3}$(MoO$_{4}$)$_{4}$. As the temperature decreases from 300 K to 5 K, a notable behavior is observed. The $g$-values for the P$_{2}$ and P$_{3}$ lines are temperature-independent, and remain constant at 2 and 2.2 above 50 K, respectively. The constant  value of $g$ indicates the paramagnetic nature of the $s$ = 1/2 spin, and corresponds to $g_{\perp}$ and $g_{\parallel}$ of Cu$^{2+}$ ions in octahedral crystal field environments \cite{PhysRevMaterials.5.014411}. 
  However, below 50 K, both \textit{g}$_{\rm P2}$ and \textit{g}$_{\rm P3}$
show a significant drop in their $g$ values suggesting the development of spin-spin correlations. Additionally, we observe that the magnetic specific heat begins to increase below 50 K. Conversely, \textit{g}$_{\rm P1}$ 
 shows a pronounced increase in its \textit{g} value as the temperature decreases, especially noticeable below 50 K. This distinct behavior of the $g$-factors suggests a complex, site-specific development of spin-spin correlations at low temperatures.  Furthermore, the pronounced temperature dependence of the $g$-factor may reflect the intriguing evolution of the effective spin network, as thermally quenched magnetic interactions become relevant. A similar phenomenon has been observed in CuGeO$_{3}$ \cite{BRILL19951683}.\\ Collectively, our X-band ESR data reveal the presence of disparate spin dynamics, alluding to an energy hierarchy of subsystems. The 1D spin correlations develop around 50 K, as evident from the narrowing of the P$_{2}$ and P$_{3}$ lines. At 30 K, some Cu$^{2+}$ spins pertaining to the P$_{1}$ line are frozen within the GHz timescale. To acquire more insight of the spin dynamics, higher-frequency ESR measurements are essential.
\\
\subsection{\label{DFT} Calculated exchange interactions \protect\\ }
\begin{figure}
	\centering
	\includegraphics[width=3.5in]{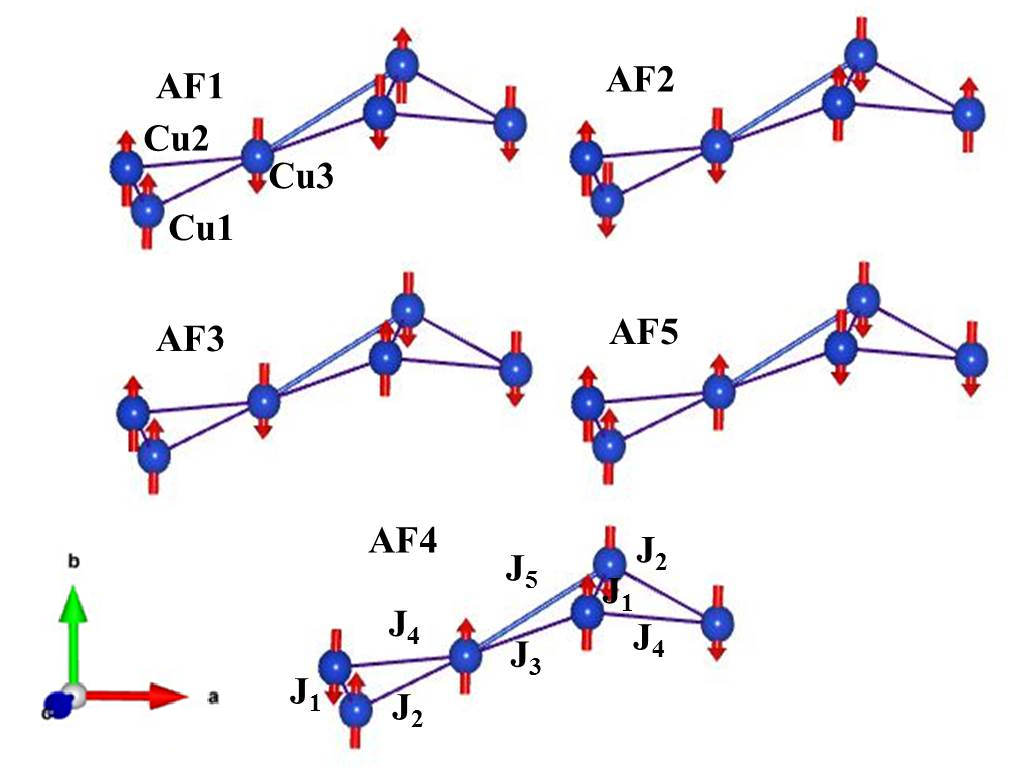}\\
	\caption{\label{fig-SS} A schematic representation of the spin arrangements for the five considered AFM spin configurations in K$_2$Cu$_3$(MoO$_4$)$_4$. In the AFM4 configuration, the exchange interactions are labeled as $J_{i}$ ($i$=1-5).}
\end{figure}
Our experimental results suggest that the compound K$_{2}$Cu$_{3}$(MoO$_{4}$)$_{4}$ exhibits multiple energy scales, each characteristic of distinct subsystems. To explore the possible magnetic spin topology and related exchange interactions, we consider two crystallographic directions: along the \( a \)-axis and  the \( c \)-axis. As illustrated in Fig.~\ref{st}(b) and \ref{st}(c), Cu$^{2+}$ ions form a distorted diamond chain along the \( a \)-axis, while along the \( c \)-axis, there are well-separated chains comprising Cu$_{2}$O$_{9}$ dimers and monomers. To determine the exchange interactions, first we consider the distorted diamond spin-chain topology (Fig.~\ref{st}(b)). In the absence of neutron diffraction experiments, no single magnetic configuration can be definitively identified. The best approach is to assume that the electronic structure remains relatively unchanged across different magnetic orders (i.e., the system maintains well-defined Cu magnetic moments) and then explore various magnetic configurations to determine the total energy differences.\\
   \begin{table}
	\caption{Calculated total energy, $\triangle E$ relative to the total energy of the FM state (E$_{\rm FM}$ = -172.8583 eV/f.u.) and magnetic moment of the Cu atom (m$_s$$^{Cu}$$\mu_{\rm B}$/atom). }
	\begin{tabular}{cccc} \hline \hline \label{afmfm}
		Configuration & $\Delta$E (meV/f.u.) & m$_s$$^{\rm Cu}$($\mu_{\rm B}$/atom)  \\ \hline
		FM& 0.0 & 0.7\\
		AFM1&$-$1.21 & 0.7  \\
		AFM2&$-$1.06& 0.7\\
		AFM3&$-$0.43& 0.7\\
		AFM4&$-$2.34& 0.7\\
		AFM5&0.31& 0.7\\ \hline
	\end{tabular}\\
\end{table}
To this end, we utilize the experimentally observed lattice parameters (see Table~\ref{table}) of K$_2$Cu$_3$(MoO$_4$)$_4$, which includes four formula units per unit cell  and consists of two chains of Cu atoms extending along the $a$-axis (Fig.~\ref{fig-SS}). Each of these chains contains 12 Cu atoms, and we have accounted for all interactions within each chain. However, we note that interactions between the two chains have not been considered in our analysis.   \\
 As a first step, we calculate the total energies for FM and various AFM configurations as shown in Fig.~\ref{fig-SS}. The system has a $C_{\rm 2h}$ point group symmetry, and the magnetic space group of the AFM configurations was identified as $\rm S2$.
 In Fig.~\ref{fig-SS}, we illustrate the exchange interactions ($J_{1}$ to $J_{5}$) between Cu atoms within a single chain.  As the two chains in the unit cell are equivalent by symmetry, we did not explicitly depict the second chain in Fig.~\ref{fig-SS}.\\
  Among these, we  find that the AFM4 configuration has lower in energy by $-$2.34 meV/f.u. relative to the FM configuration (see Table~\ref{afmfm}). Therefore, the magnetic ground state of K$_2$Cu$_3$(MoO$_4$)$_4$ is AFM4, as depicted in Fig.~\ref{fig-SS} (only the top layer is shown). The magnetic moment of Cu ions is calculated as 0.7 $\mu_{\textrm{B}}$/atom, which is slightly  smaller than the expected ordered moment 1 $\mu_{\rm B}$ for a Cu$^{2+}$ ion \cite{PhysRevB.83.104401}. This discrepancy may be due to the fact that DFT-based calculations typically underestimate the moment size for 3$d$ transition metals by transferring some spin density onto the surrounding O$^{2-}$ ions. Moreover, this discrepancy could arise from quantum entanglement and spin fluctuations in $s = 1/2$ systems. We note that the experimentally determined magnetic moment of azurite at low temperatures is 0.6 $\mu_{B}$ \cite{PhysRevB.83.104401}.  \\
 To figure out the nature and magnitude of exchange-interaction parameters, we choose the Heisenberg model. The Heisenberg Hamiltonian can be expressed as $\mathcal{H}$ = E$_0$ $-$ $\Sigma$$_{i,j}$$J_{ij}$$S$$_i$$S$$_j$, where E$_0$ is the total energy for all the spin-independent  interactions, $J_{ij}$ is the exchange interaction parameters between the Cu atoms at sites  \textit{i} and  \textit{j}, $S$$_i$ and $S$$_j$ are the unit vectors representing the directions of the local magnetic moments at the sites \textit{i} and \textit{j}, respectively. Here, $J$ $<$ 0 implies AFM and $J$ $>$ 0 indicates FM interactions. By solving the set of equations,
\begin{align*}
E_{\text{FM}} &= E_0 - 4J_1 - 4J_2 - 2J_3 - 4J_4 - 2J_5, \\
E_{\text{AF1}} &= E_0 + 4J_2 - 2J_3 + 2J_5, \\
E_{\text{AF2}} &= E_0 + 4J_1 + 2J_3, \\
E_{\text{AF3}} &= E_0 + 2J_3 + 4J_4 - 2J_5, \\
E_{\text{AF4}} &= E_0 + 4J_1 - 4J_2 - 2J_3 + 4J_4 + 2J_5, \\
E_{\text{AF5}} &= E_0 - 4J_1 - 4J_2 + 2J_3 - 4J_4 + 2J_5.
\end{align*}
   We find the values of $J_{ij}$ as listed in Table~\ref{listJ}. The calculated $J_1$, $J_4$, and $J_5$ are found to be AFM \begin{figure}
   	\centering
   	\includegraphics[width=0.4\textwidth]{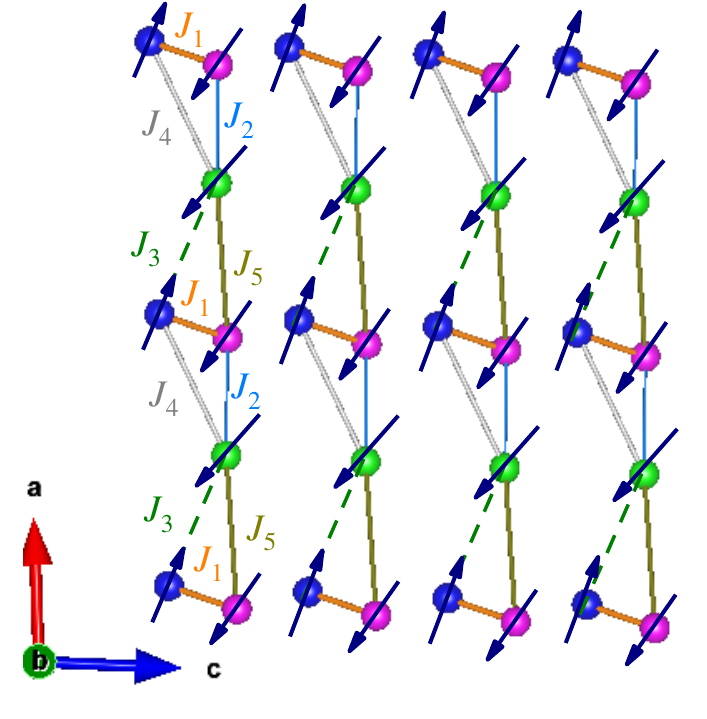}\\
   	\caption{\label{LDA}Depicts the resulting spin model derived from DFT + $U_{\rm eff}$ (GGA) calculations for K$_2$Cu$_3$(MoO$_4$)$_4$ using  VASP. It consists of an infinite zig-zag distorted diamond chain along the \( a \)-direction, characterized by \( J_1 \) intra-dimer interactions and \( J_2 \), \( J_3 \), \( J_4 \), and \( J_5 \) interactions between dimers and monomers along the \( a \)-axis, collectively forming a diamond lattice.
   	}
   \end{figure}
   \begin{figure}
   	\centering
   	\includegraphics[width=0.44\textwidth]{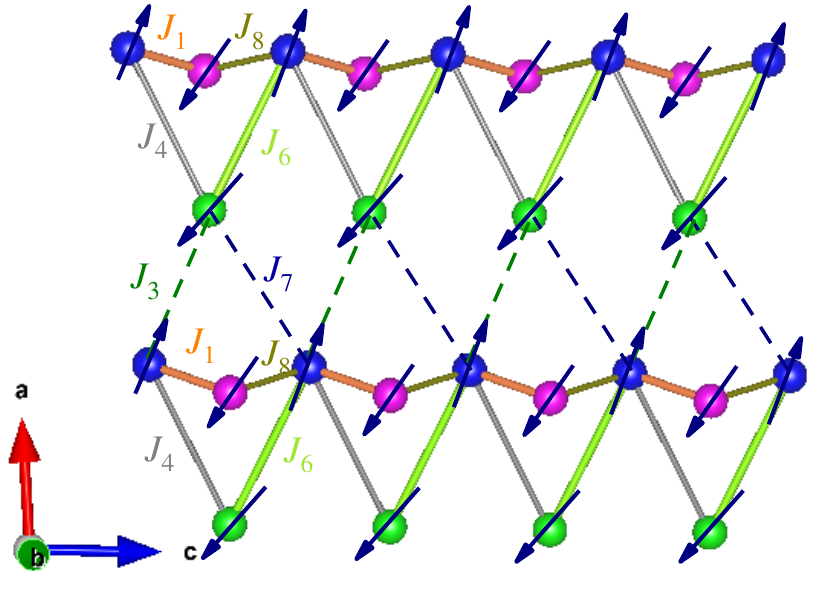}\\
   	\caption{\label{LDAM}Schematic representation of the spin model obtained from DFT + \( U \) (LDA) calculations for K\(_2\)Cu\(_3\)(MoO\(_4\))\(_4\) using the OpenMX code. It features a chain of Cu\(_2\)O\(_9\) dimers along the \( c \)-axis, with intra-dimer interaction \( J_8 \) and inter-chain interactions \( J_4 \), \( J_6 \), \( J_3 \), and \( J_7 \), collectively forming a coupled zig-zag chain of dimers along the \( c \)-axis.
   	}
   \end{figure}
   \begin{table}
	\caption{Heisenberg exchange coupling constants ($J_{ij}$ in K) calculated using the DFT + $U_{\rm eff}$ (GGA) approach, along with the corresponding distances between the Cu atoms. }
	\begin{tabular}{ccccc} \hline \hline \label{listJ}
		$J_{ij}$ & Cu-Cu (\AA) & K (U$_{\rm eff}$ = 3.6 eV) \\ \hline
		$J_{1}$ & 3.11 & $-$1.9 \\
		$J_{2}$ & 4.79 & 0.3 \\
		$J_{3}$ & 5.54 & 2.1  \\
		$J_{4}$ & 5.78 & $-$0.8\\
		$J_{5}$ & 6.62 & $-$1.2  \\ 
		\hline
	\end{tabular}\\
\end{table}
\begin{table}[b]
	\caption{Estimated Heisenberg exchange coupling constants using DFT + $U$ (LDA) approach ($J_{ij}$ in K) and the corresponding distances between the Cu atoms. }
	\begin{tabular}{ccccc} \hline \hline\label{listJlda}
		$J_{ij}^{\rm LDA}$ & Cu-Cu (\AA) & K ($U$ = 2 eV)& K ($U$ =3 eV)& K ($U$ =4 eV)  \\ \hline
		$J_{1}$ & 3.11 & $-$8.816& $-$ 6.264&$-$4.176  \\
		$J_{3}$ & 5.54& $-$1.39&$-$1.044 & $-$0.812 \\
		$J_{4}$ & 5.78 &0.232&0.116 & 0  \\
		$J_{6}$ & 5.67&$-$7.88 &$-$5.8& $-$4.29  \\
		$J_{7}$ & 6.13 &$-$1.74 & $-$1.276 &$-$0.928  \\
		$J_{8}$ & 3.15 &$-$2.784&$-$1.508 &$-$0.232  \\ \hline
	\end{tabular}\\
\end{table}
 while $J_2$ and $J_3$ are FM. Here, $J_1$ and $J_5$ correspond to the shortest and longest distance between the Cu atoms, respectively. The exchange interactions presented in Table~\ref{listJ} reveal that $J_1$ significantly surpasses $J_2$ and $J_4$, indicating that the K$_2$Cu$_3$(MoO$_4$)$_4$ system can be characterized as a distorted diamond chain with a periodic array of $J_1$ dimers and competing $J_2$, $J_4$, $J_3$, and $J_5$ monomer spins (see Fig.~\ref{LDA}). The calculated Weiss temperature ($\theta_{\rm CW}^{\rm GGA}$) is $-$3.9 K, aligning with the experimental Weiss temperature ($\theta_{\rm CW}^{\rm Ex}$ = $-$3.82 K) derived from the low-temperature 1/$\chi(T)$ plot (Fig.~\ref{chi}(b)).  Despite the presence of significant competing exchange interactions, the system is not highly frustrated. This moderate frustration facilitates the achievement of long-range AFM order through the alternating spin directions of the monomers along the chain. It is worth to note that Fig.~\ref{LDA} and Fig.~\ref{LDAM} provide a schematic representation of the proposed AFM configuration, where the arrows depict the spins of individual atoms. These arrows are chosen to illustrate the approximate nature of exchange interactions and their corresponding interaction strengths ($J_{1}$, $J_{2}$, $J_{3}$, etc,.). However, they do not represent the actual spin orientations. Future neutron diffraction experiments may be required to determine the precise spin orientations. \\
Up to now, our theoretical calculations effectively reproduce the estimated low-temperature Weiss temperature by predicting a possible spin structure for the distorted diamond chain along the $a$-axis. However, it is worth noting that
 the DFT calculations discussed above failed to capture the higher-energy scale of spin-singlet correlations around 20 K. This is probably because the GGA $+$ $U_{\rm eff}$ approach primarily emphasizes the diamond chain structure along the $a$-axis, accounting exclusively for the Cu$_{2}$O$_{9}$ dimer configuration associated with this direction. Consequently, this simplification may overlook additional exchange pathways, particularly neglecting longer-range exchange paths. To address this limitation in the DFT $+$ U$_{\rm eff}$ approach, which assumes the $\Gamma$-point of the Brillouin zone, we extend our calculations (see below) using  magnetic-force linear response theory, which inherently accounts for long-range magnetic exchange interactions \cite{LIECHTENSTEIN198765}. 
\\ To incorporate both the intra-dimer interactions along the $c$-axis and the additional interactions between dimers mediated by monomers along the 
$a$-axis, we employed DFT$+$ $U$ calculations using the LDA exchange-correlation functional within the OpenMX framework for three different values of $U$ (see Table~\ref{listJlda}). The LDA$+$$U$ approach highlights the $c$-axis configuration (see Fig.~\ref{LDAM}), featuring a chain of Cu$_{2}$O$_{9}$ dimers coupled through AFM interactions $J_{1}$ and intra-dimer interactions $J_{8}$ (dark yellow bond in Fig.~\ref{LDAM}). Additionally, it identifies two AFM interactions between dimers and monomers, $J_6$ and $J_7$ (Fig.~\ref{LDAM}). Interestingly, it is noted that the sum of all exchange interactions obtained from the spin model in Fig.~\ref{LDAM} for $U$ = 4 eV results in $J/k_{\rm B}$ = 10.43 K. The calculated value of $|\theta_{\rm CW}| = 5.21$ K form $J/k_{\rm B} = 10.43$ K is slightly higher than the low-temperature Curie-Weiss temperature. This suggests that even we consider long-range magnetic exchange interactions our second approach also fails to accurately reproduce the experimental results, including subdominant ferromagnetic interactions at low-temperatures and high-temperature energy scale.  \\
Furthermore, we clarify that two distinct DFT-based methods were employed to calculate the magnetic exchange
	interactions.  The discrepancies in the value of exchange parameters obtained from the two approaches can be attributed to the intrinsic differences between these functionals  as well as the distinct approaches used to extract the magnetic exchange interactions. Admittedly, neither approach fully captures the magnetic behavior of the studied compound. However, both DFT calculations consistently reveal the presence of dominant intra-dimer interactions, despite differences in the underlying spin topology. 
\section{Conclusions}
In summary, we have investigated the crystal structure and ground-state properties of a $s$ = 1/2 compound K$_2$Cu$_3$(MoO$_4$)$_4$ through diverse experimental techniques of SXRD, magnetic susceptibility, specific heat, and ESR, complemented by DFT calculations. Rietveld refinement of  SXRD data suggest that along the $a$-axis, Cu$^{2+}$ ions are arranged in a triangular pattern with alternating dimers and monomers, forming a coupled, distorted diamond chain. On the other hand, along the $c$-axis, Cu$^{2+}$ ions form a zig-zag chain of Cu$_{2}$O$_{9}$ dimers interspersed with a uniform chain of monomers. \\Magnetic susceptibility,  magnetic specfiic heat, and ESR reveal two magnetic transitions: a spin-singlet state at 20 K associated with Cu$_{2}$O$_{9}$ dimers  and short-range correlations at 3.68 K between dimers and monomers. 
Furthermore, the temperature dependence of inverse magnetic susceptibility data indicate the coexistence of ferromagnetic and AFM interactions between Cu$^{2+}$
moments at high and low temperatures, respectively, in  K$_2$Cu$_3$(MoO$_4$)$_4$.  
  \\ A $\lambda$-type anomaly in 
specific heat data unambiguously demonstrate the presence of long-range AFM order at $T_{\rm N}$ = 2.3 K, which is consistent with the magnetic susceptibility measurements. A comprehensive field-temperature phase diagram, constructed from isothermal magnetization and specific heat data, unveils a sequence of field-induced phase transitions. These include a field-induced phase AFM2, and the 1/3 plateau phase. The appearance of a 1/3 magnetization plateau phase above 3.5 T, characterized by a saturation value of 
0.35 $\mu_{\rm B}$, points to the polarization of monomer spins.  \\ Our ESR experiments reveal distinct magnetic sites and $g$-factor anisotropy. Three different linewidths highlight disparate spin dynamics, suggesting an energy hierarchy of subsystems consistent with magnetic susceptibility features. A broad maximum around 30 K in the ESR linewidth provides microscopic evidence of a spin-singlet state. Two different DFT calculations suggest the presence of dominant dimer interactions; however, precisely determining the spin topology remains challenging due to the complexity introduced by numerous long-range exchange pathways, calling for future investigations. \\
\section{ACKNOWLEDGMENTS}
G. S. M. gratefully acknowledges Professor F. C. Chou for his invaluable suggestions. The work at
SKKU was supported by the National Research Foundation (NRF) of Korea (Grants No. RS-2023-00209121 and
No. 2020R1A5A1016518).
R. S. acknowledges the financial support provided by the Ministry of Science and Technology in Taiwan under Project No. NSTC-113-2124-M-001-003 and No. NSTC-113-2112M001-045-MY3,  as well as support from Academia Sinica for the budget of AS-iMATE11312. financial support from the Center of Atomic Initiative for New Materials (AIMat), National Taiwan University,  under Project No. 113L900801. R.K.U. would like to acknowledge the IITR for the Faculty Initiation Grant (FIG-101068). I.P.M thanks the Banaras Hindu University for the Institutes of Eminence (IoE) Seed Grant. I.P.M also acknowledges the funding by the UGC-DAE Consortium for Scientific Research, Mumbai, India (Project No. CRS/2022-23/03/863), and the Department of Science and Technology-Science and Engineering Research Board (DST-SERB), India through Research Grants. E.M. is supported by the Canada Research Chairs program, the Natural Science and Engineering Research Council of Canada, and the Canadian Foundation for Innovation. C.L.H.’s work is supported by the
National Science and Technology Council in Taiwan with a
Grant No. NSTC 113-2112-M-006-027. 
\section{Data availability}
The data that support the findings of this study are available at Ref. \cite{data}.
\bibliography{kcmo}

\begin{thebibliography}{70}%
\makeatletter
\providecommand \@ifxundefined [1]{%
 \@ifx{#1\undefined}
}%
\providecommand \@ifnum [1]{%
 \ifnum #1\expandafter \@firstoftwo
 \else \expandafter \@secondoftwo
 \fi
}%
\providecommand \@ifx [1]{%
 \ifx #1\expandafter \@firstoftwo
 \else \expandafter \@secondoftwo
 \fi
}%
\providecommand \natexlab [1]{#1}%
\providecommand \enquote  [1]{``#1''}%
\providecommand \bibnamefont  [1]{#1}%
\providecommand \bibfnamefont [1]{#1}%
\providecommand \citenamefont [1]{#1}%
\providecommand \href@noop [0]{\@secondoftwo}%
\providecommand \href [0]{\begingroup \@sanitize@url \@href}%
\providecommand \@href[1]{\@@startlink{#1}\@@href}%
\providecommand \@@href[1]{\endgroup#1\@@endlink}%
\providecommand \@sanitize@url [0]{\catcode `\\12\catcode `\$12\catcode
  `\&12\catcode `\#12\catcode `\^12\catcode `\_12\catcode `\%12\relax}%
\providecommand \@@startlink[1]{}%
\providecommand \@@endlink[0]{}%
\providecommand \url  [0]{\begingroup\@sanitize@url \@url }%
\providecommand \@url [1]{\endgroup\@href {#1}{\urlprefix }}%
\providecommand \urlprefix  [0]{URL }%
\providecommand \Eprint [0]{\href }%
\providecommand \doibase [0]{https://doi.org/}%
\providecommand \selectlanguage [0]{\@gobble}%
\providecommand \bibinfo  [0]{\@secondoftwo}%
\providecommand \bibfield  [0]{\@secondoftwo}%
\providecommand \translation [1]{[#1]}%
\providecommand \BibitemOpen [0]{}%
\providecommand \bibitemStop [0]{}%
\providecommand \bibitemNoStop [0]{.\EOS\space}%
\providecommand \EOS [0]{\spacefactor3000\relax}%
\providecommand \BibitemShut  [1]{\csname bibitem#1\endcsname}%
\let\auto@bib@innerbib\@empty
\bibitem [{\citenamefont {Anderson}(1973)}]{ANDERSON1973153}%
  \BibitemOpen
  \bibfield  {author} {\bibinfo {author} {\bibfnamefont {P.~W.}\ \bibnamefont
  {Anderson}},\ }\bibfield  {title} {\bibinfo {title} {Resonating valence
  bonds: A new kind of insulator?},\ }\href
  {https://doi.org/https://doi.org/10.1016/0025-5408(73)90167-0} {\bibfield
  {journal} {\bibinfo  {journal} {Mater. Res. Bull.}\ }\textbf {\bibinfo
  {volume} {8}},\ \bibinfo {pages} {153 } (\bibinfo {year} {1973})}\BibitemShut
  {NoStop}%
\bibitem [{\citenamefont {Matan}\ \emph {et~al.}(2010)\citenamefont {Matan},
  \citenamefont {Ono}, \citenamefont {Fukumoto}, \citenamefont {Sato},
  \citenamefont {Yamaura}, \citenamefont {Yano}, \citenamefont {Morita},\ and\
  \citenamefont {Tanaka}}]{Matan2010}%
  \BibitemOpen
  \bibfield  {author} {\bibinfo {author} {\bibfnamefont {K.}~\bibnamefont
  {Matan}}, \bibinfo {author} {\bibfnamefont {T.}~\bibnamefont {Ono}}, \bibinfo
  {author} {\bibfnamefont {Y.}~\bibnamefont {Fukumoto}}, \bibinfo {author}
  {\bibfnamefont {T.~J.}\ \bibnamefont {Sato}}, \bibinfo {author}
  {\bibfnamefont {J.}~\bibnamefont {Yamaura}}, \bibinfo {author} {\bibfnamefont
  {M.}~\bibnamefont {Yano}}, \bibinfo {author} {\bibfnamefont {K.}~\bibnamefont
  {Morita}},\ and\ \bibinfo {author} {\bibfnamefont {H.}~\bibnamefont
  {Tanaka}},\ }\bibfield  {title} {\bibinfo {title} {Pinwheel valence-bond
  solid and triplet excitations in the two-dimensional deformed kagome
  lattice},\ }\href {https://doi.org/10.1038/nphys1761} {\bibfield  {journal}
  {\bibinfo  {journal} {Nat. Phys.}\ }\textbf {\bibinfo {volume} {6}},\
  \bibinfo {pages} {865} (\bibinfo {year} {2010})}\BibitemShut {NoStop}%
\bibitem [{\citenamefont {Balents}(2010)}]{Balents2010}%
  \BibitemOpen
  \bibfield  {author} {\bibinfo {author} {\bibfnamefont {L.}~\bibnamefont
  {Balents}},\ }\bibfield  {title} {\bibinfo {title} {Spin liquids in
  frustrated magnets},\ }\href {https://doi.org/10.1038/nature08917} {\bibfield
   {journal} {\bibinfo  {journal} {Nature}\ }\textbf {\bibinfo {volume}
  {464}},\ \bibinfo {pages} {199} (\bibinfo {year} {2010})}\BibitemShut
  {NoStop}%
\bibitem [{\citenamefont {Zhou}\ \emph {et~al.}(2017)\citenamefont {Zhou},
  \citenamefont {Kanoda},\ and\ \citenamefont {Ng}}]{RevModPhys.89.025003}%
  \BibitemOpen
  \bibfield  {author} {\bibinfo {author} {\bibfnamefont {Y.}~\bibnamefont
  {Zhou}}, \bibinfo {author} {\bibfnamefont {K.}~\bibnamefont {Kanoda}},\ and\
  \bibinfo {author} {\bibfnamefont {T.-K.}\ \bibnamefont {Ng}},\ }\bibfield
  {title} {\bibinfo {title} {Quantum spin liquid states},\ }\href
  {https://doi.org/10.1103/RevModPhys.89.025003} {\bibfield  {journal}
  {\bibinfo  {journal} {Rev. Mod. Phys.}\ }\textbf {\bibinfo {volume} {89}},\
  \bibinfo {pages} {025003} (\bibinfo {year} {2017})}\BibitemShut {NoStop}%
\bibitem [{\citenamefont {Savary}\ and\ \citenamefont
  {Balents}(2016)}]{Savary_2016}%
  \BibitemOpen
  \bibfield  {author} {\bibinfo {author} {\bibfnamefont {L.}~\bibnamefont
  {Savary}}\ and\ \bibinfo {author} {\bibfnamefont {L.}~\bibnamefont
  {Balents}},\ }\bibfield  {title} {\bibinfo {title} {Quantum spin liquids: a
  review},\ }\href {https://doi.org/10.1088/0034-4885/80/1/016502} {\bibfield
  {journal} {\bibinfo  {journal} {Rep. Prog. Phys.}\ }\textbf {\bibinfo
  {volume} {80}},\ \bibinfo {pages} {016502} (\bibinfo {year}
  {2016})}\BibitemShut {NoStop}%
\bibitem [{\citenamefont {Broholm}\ \emph {et~al.}(2020)\citenamefont
  {Broholm}, \citenamefont {Cava}, \citenamefont {Kivelson}, \citenamefont
  {Nocera}, \citenamefont {Norman},\ and\ \citenamefont
  {Senthil}}]{doi:10.1126/science.aay0668}%
  \BibitemOpen
  \bibfield  {author} {\bibinfo {author} {\bibfnamefont {C.}~\bibnamefont
  {Broholm}}, \bibinfo {author} {\bibfnamefont {R.~J.}\ \bibnamefont {Cava}},
  \bibinfo {author} {\bibfnamefont {S.~A.}\ \bibnamefont {Kivelson}}, \bibinfo
  {author} {\bibfnamefont {D.~G.}\ \bibnamefont {Nocera}}, \bibinfo {author}
  {\bibfnamefont {M.~R.}\ \bibnamefont {Norman}},\ and\ \bibinfo {author}
  {\bibfnamefont {T.}~\bibnamefont {Senthil}},\ }\bibfield  {title} {\bibinfo
  {title} {Quantum spin liquids},\ }\href
  {https://doi.org/10.1126/science.aay0668} {\bibfield  {journal} {\bibinfo
  {journal} {Science}\ }\textbf {\bibinfo {volume} {367}},\ \bibinfo {pages}
  {eaay0668} (\bibinfo {year} {2020})}\BibitemShut {NoStop}%
\bibitem [{\citenamefont {Jeon}\ \emph {et~al.}(2024)\citenamefont {Jeon},
  \citenamefont {Wulferding}, \citenamefont {Choi}, \citenamefont {Lee},
  \citenamefont {Nam}, \citenamefont {Kim}, \citenamefont {Lee}, \citenamefont
  {Jang}, \citenamefont {Park}, \citenamefont {Lee}, \citenamefont {Choi},
  \citenamefont {Lee}, \citenamefont {Nojiri},\ and\ \citenamefont
  {Choi}}]{Jeon2024}%
  \BibitemOpen
  \bibfield  {author} {\bibinfo {author} {\bibfnamefont {S.}~\bibnamefont
  {Jeon}}, \bibinfo {author} {\bibfnamefont {D.}~\bibnamefont {Wulferding}},
  \bibinfo {author} {\bibfnamefont {Y.}~\bibnamefont {Choi}}, \bibinfo {author}
  {\bibfnamefont {S.}~\bibnamefont {Lee}}, \bibinfo {author} {\bibfnamefont
  {K.}~\bibnamefont {Nam}}, \bibinfo {author} {\bibfnamefont {K.~H.}\
  \bibnamefont {Kim}}, \bibinfo {author} {\bibfnamefont {M.}~\bibnamefont
  {Lee}}, \bibinfo {author} {\bibfnamefont {T.-H.}\ \bibnamefont {Jang}},
  \bibinfo {author} {\bibfnamefont {J.-H.}\ \bibnamefont {Park}}, \bibinfo
  {author} {\bibfnamefont {S.}~\bibnamefont {Lee}}, \bibinfo {author}
  {\bibfnamefont {S.}~\bibnamefont {Choi}}, \bibinfo {author} {\bibfnamefont
  {C.}~\bibnamefont {Lee}}, \bibinfo {author} {\bibfnamefont {H.}~\bibnamefont
  {Nojiri}},\ and\ \bibinfo {author} {\bibfnamefont {K.-Y.}\ \bibnamefont
  {Choi}},\ }\bibfield  {title} {\bibinfo {title} {One-ninth magnetization
  plateau stabilized by spin entanglement in a kagome antiferromagnet},\ }\href
  {https://doi.org/10.1038/s41567-023-02318-7} {\bibfield  {journal} {\bibinfo
  {journal} {Nat. Phys.}\ }\textbf {\bibinfo {volume} {20}},\ \bibinfo {pages}
  {435} (\bibinfo {year} {2024})}\BibitemShut {NoStop}%
\bibitem [{\citenamefont {Khatua}\ \emph {et~al.}(2023)\citenamefont {Khatua},
  \citenamefont {Sana}, \citenamefont {Zorko}, \citenamefont {Gomilšek},
  \citenamefont {Sethupathi}, \citenamefont {Rao}, \citenamefont {Baenitz},
  \citenamefont {Schmidt},\ and\ \citenamefont {Khuntia}}]{KHATUA20231}%
  \BibitemOpen
  \bibfield  {author} {\bibinfo {author} {\bibfnamefont {J.}~\bibnamefont
  {Khatua}}, \bibinfo {author} {\bibfnamefont {B.}~\bibnamefont {Sana}},
  \bibinfo {author} {\bibfnamefont {A.}~\bibnamefont {Zorko}}, \bibinfo
  {author} {\bibfnamefont {M.}~\bibnamefont {Gomilšek}}, \bibinfo {author}
  {\bibfnamefont {K.}~\bibnamefont {Sethupathi}}, \bibinfo {author}
  {\bibfnamefont {M.~R.}\ \bibnamefont {Rao}}, \bibinfo {author} {\bibfnamefont
  {M.}~\bibnamefont {Baenitz}}, \bibinfo {author} {\bibfnamefont
  {B.}~\bibnamefont {Schmidt}},\ and\ \bibinfo {author} {\bibfnamefont
  {P.}~\bibnamefont {Khuntia}},\ }\bibfield  {title} {\bibinfo {title}
  {Experimental signatures of quantum and topological states in frustrated
  magnetism},\ }\href
  {https://doi.org/https://doi.org/10.1016/j.physrep.2023.09.008} {\bibfield
  {journal} {\bibinfo  {journal} {Phys. Rep.}\ }\textbf {\bibinfo {volume}
  {1041}},\ \bibinfo {pages} {1} (\bibinfo {year} {2023})}\BibitemShut
  {NoStop}%
\bibitem [{\citenamefont {Kikuchi}\ \emph {et~al.}(2004)\citenamefont
  {Kikuchi}, \citenamefont {Fujii}, \citenamefont {Chiba}, \citenamefont
  {Mitsudo}, \citenamefont {Idehara},\ and\ \citenamefont
  {Kuwai}}]{KIKUCHI2004900}%
  \BibitemOpen
  \bibfield  {author} {\bibinfo {author} {\bibfnamefont {H.}~\bibnamefont
  {Kikuchi}}, \bibinfo {author} {\bibfnamefont {Y.}~\bibnamefont {Fujii}},
  \bibinfo {author} {\bibfnamefont {M.}~\bibnamefont {Chiba}}, \bibinfo
  {author} {\bibfnamefont {S.}~\bibnamefont {Mitsudo}}, \bibinfo {author}
  {\bibfnamefont {T.}~\bibnamefont {Idehara}},\ and\ \bibinfo {author}
  {\bibfnamefont {T.}~\bibnamefont {Kuwai}},\ }\bibfield  {title} {\bibinfo
  {title} {Experimental evidence of the one-third magnetization plateau in the
  diamond chain compound
  \text{Cu$_{3}$}\text{(CO$_{3}$)$_{2}$}\text{(OH)$_{2}$}},\ }\href
  {https://doi.org/https://doi.org/10.1016/j.jmmm.2003.12.619} {\bibfield
  {journal} {\bibinfo  {journal} {J. Magn. Magn. Mater.}\ }\textbf {\bibinfo
  {volume} {272-276}},\ \bibinfo {pages} {900} (\bibinfo {year} {2004})},\
  \bibinfo {note} {proceedings of the International Conference on Magnetism
  (ICM 2003)}\BibitemShut {NoStop}%
\bibitem [{\citenamefont {Kikuchi}\ \emph {et~al.}(2005)\citenamefont
  {Kikuchi}, \citenamefont {Fujii}, \citenamefont {Chiba}, \citenamefont
  {Mitsudo}, \citenamefont {Idehara}, \citenamefont {Tonegawa}, \citenamefont
  {Okamoto}, \citenamefont {Sakai}, \citenamefont {Kuwai},\ and\ \citenamefont
  {Ohta}}]{PhysRevLett.94.227201}%
  \BibitemOpen
  \bibfield  {author} {\bibinfo {author} {\bibfnamefont {H.}~\bibnamefont
  {Kikuchi}}, \bibinfo {author} {\bibfnamefont {Y.}~\bibnamefont {Fujii}},
  \bibinfo {author} {\bibfnamefont {M.}~\bibnamefont {Chiba}}, \bibinfo
  {author} {\bibfnamefont {S.}~\bibnamefont {Mitsudo}}, \bibinfo {author}
  {\bibfnamefont {T.}~\bibnamefont {Idehara}}, \bibinfo {author} {\bibfnamefont
  {T.}~\bibnamefont {Tonegawa}}, \bibinfo {author} {\bibfnamefont
  {K.}~\bibnamefont {Okamoto}}, \bibinfo {author} {\bibfnamefont
  {T.}~\bibnamefont {Sakai}}, \bibinfo {author} {\bibfnamefont
  {T.}~\bibnamefont {Kuwai}},\ and\ \bibinfo {author} {\bibfnamefont
  {H.}~\bibnamefont {Ohta}},\ }\bibfield  {title} {\bibinfo {title}
  {Experimental observation of the $1/3$ magnetization plateau in the
  diamond-chain compound
  \text{Cu$_{3}$}\text{(CO$_{3}$)$_{2}$}\text{(OH)$_{2}$}},\ }\href
  {https://doi.org/10.1103/PhysRevLett.94.227201} {\bibfield  {journal}
  {\bibinfo  {journal} {Phys. Rev. Lett.}\ }\textbf {\bibinfo {volume} {94}},\
  \bibinfo {pages} {227201} (\bibinfo {year} {2005})}\BibitemShut {NoStop}%
\bibitem [{\citenamefont {Spence}\ and\ \citenamefont
  {Ewing}(1958)}]{PhysRev.112.1544}%
  \BibitemOpen
  \bibfield  {author} {\bibinfo {author} {\bibfnamefont {R.~D.}\ \bibnamefont
  {Spence}}\ and\ \bibinfo {author} {\bibfnamefont {R.~D.}\ \bibnamefont
  {Ewing}},\ }\bibfield  {title} {\bibinfo {title} {Evidence for
  antiferromagnetism in
  \text{Cu$_{3}$}\text{(CO$_{3}$)$_{2}$}\text{(OH)$_{2}$}},\ }\href
  {https://doi.org/10.1103/PhysRev.112.1544} {\bibfield  {journal} {\bibinfo
  {journal} {Phys. Rev.}\ }\textbf {\bibinfo {volume} {112}},\ \bibinfo {pages}
  {1544} (\bibinfo {year} {1958})}\BibitemShut {NoStop}%
\bibitem [{\citenamefont {Forstat}\ \emph {et~al.}(1959)\citenamefont
  {Forstat}, \citenamefont {Taylor},\ and\ \citenamefont
  {King}}]{10.1063/1.1730553}%
  \BibitemOpen
  \bibfield  {author} {\bibinfo {author} {\bibfnamefont {H.}~\bibnamefont
  {Forstat}}, \bibinfo {author} {\bibfnamefont {G.}~\bibnamefont {Taylor}},\
  and\ \bibinfo {author} {\bibfnamefont {B.~R.}\ \bibnamefont {King}},\
  }\bibfield  {title} {\bibinfo {title} {{Low‐Temperature Heat Capacity of
  Azurite}},\ }\href {https://doi.org/10.1063/1.1730553} {\bibfield  {journal}
  {\bibinfo  {journal} {J. Chem. Phys.}\ }\textbf {\bibinfo {volume} {31}},\
  \bibinfo {pages} {929} (\bibinfo {year} {1959})}\BibitemShut {NoStop}%
\bibitem [{\citenamefont {{van der Lugt}}\ and\ \citenamefont
  {Poulis}(1959)}]{VANDERLUGT19591313}%
  \BibitemOpen
  \bibfield  {author} {\bibinfo {author} {\bibfnamefont {W.}~\bibnamefont {{van
  der Lugt}}}\ and\ \bibinfo {author} {\bibfnamefont {N.}~\bibnamefont
  {Poulis}},\ }\bibfield  {title} {\bibinfo {title} {Proton magnetic resonance
  in azurite},\ }\href
  {https://doi.org/https://doi.org/10.1016/0031-8914(59)90054-0} {\bibfield
  {journal} {\bibinfo  {journal} {Physica}\ }\textbf {\bibinfo {volume} {25}},\
  \bibinfo {pages} {1313} (\bibinfo {year} {1959})}\BibitemShut {NoStop}%
\bibitem [{\citenamefont {Frikkee}\ and\ \citenamefont {{Van Den
  handel}}(1962)}]{FRIKKEE1962269}%
  \BibitemOpen
  \bibfield  {author} {\bibinfo {author} {\bibfnamefont {E.}~\bibnamefont
  {Frikkee}}\ and\ \bibinfo {author} {\bibfnamefont {J.}~\bibnamefont {{Van Den
  handel}}},\ }\bibfield  {title} {\bibinfo {title} {The magnetic behaviour of
  azurite},\ }\href
  {https://doi.org/https://doi.org/10.1016/0031-8914(62)90046-0} {\bibfield
  {journal} {\bibinfo  {journal} {Physica}\ }\textbf {\bibinfo {volume} {28}},\
  \bibinfo {pages} {269} (\bibinfo {year} {1962})}\BibitemShut {NoStop}%
\bibitem [{\citenamefont {Love}\ \emph {et~al.}(1970)\citenamefont {Love},
  \citenamefont {{Duncan 0999 09}}, \citenamefont {Bailey},\ and\ \citenamefont
  {Forstat}}]{LOVE1970290}%
  \BibitemOpen
  \bibfield  {author} {\bibinfo {author} {\bibfnamefont {N.}~\bibnamefont
  {Love}}, \bibinfo {author} {\bibfnamefont {T.}~\bibnamefont {{Duncan 0999
  09}}}, \bibinfo {author} {\bibfnamefont {P.}~\bibnamefont {Bailey}},\ and\
  \bibinfo {author} {\bibfnamefont {H.}~\bibnamefont {Forstat}},\ }\bibfield
  {title} {\bibinfo {title} {Spin-flopping in azurite},\ }\href
  {https://doi.org/https://doi.org/10.1016/0375-9601(70)90144-1} {\bibfield
  {journal} {\bibinfo  {journal} {Physics Letters A}\ }\textbf {\bibinfo
  {volume} {33}},\ \bibinfo {pages} {290} (\bibinfo {year} {1970})}\BibitemShut
  {NoStop}%
\bibitem [{\citenamefont {Gu}\ and\ \citenamefont
  {Su}(2006)}]{PhysRevLett.97.089701}%
  \BibitemOpen
  \bibfield  {author} {\bibinfo {author} {\bibfnamefont {B.}~\bibnamefont
  {Gu}}\ and\ \bibinfo {author} {\bibfnamefont {G.}~\bibnamefont {Su}},\
  }\bibfield  {title} {\bibinfo {title} {Comment on ``experimental observation
  of the $1/3$ magnetization plateau in the diamond-chain compound
  \text{Cu$_{3}$}\text{(CO$_{3}$)$_{2}$}\text{(OH)$_{2}$}''},\ }\href
  {https://doi.org/10.1103/PhysRevLett.97.089701} {\bibfield  {journal}
  {\bibinfo  {journal} {Phys. Rev. Lett.}\ }\textbf {\bibinfo {volume} {97}},\
  \bibinfo {pages} {089701} (\bibinfo {year} {2006})}\BibitemShut {NoStop}%
\bibitem [{\citenamefont {Rule}\ \emph {et~al.}(2008)\citenamefont {Rule},
  \citenamefont {Wolter}, \citenamefont {S\"ullow}, \citenamefont {Tennant},
  \citenamefont {Br\"uhl}, \citenamefont {K\"ohler}, \citenamefont {Wolf},
  \citenamefont {Lang},\ and\ \citenamefont
  {Schreuer}}]{PhysRevLett.100.117202}%
  \BibitemOpen
  \bibfield  {author} {\bibinfo {author} {\bibfnamefont {K.~C.}\ \bibnamefont
  {Rule}}, \bibinfo {author} {\bibfnamefont {A.~U.~B.}\ \bibnamefont {Wolter}},
  \bibinfo {author} {\bibfnamefont {S.}~\bibnamefont {S\"ullow}}, \bibinfo
  {author} {\bibfnamefont {D.~A.}\ \bibnamefont {Tennant}}, \bibinfo {author}
  {\bibfnamefont {A.}~\bibnamefont {Br\"uhl}}, \bibinfo {author} {\bibfnamefont
  {S.}~\bibnamefont {K\"ohler}}, \bibinfo {author} {\bibfnamefont
  {B.}~\bibnamefont {Wolf}}, \bibinfo {author} {\bibfnamefont {M.}~\bibnamefont
  {Lang}},\ and\ \bibinfo {author} {\bibfnamefont {J.}~\bibnamefont
  {Schreuer}},\ }\bibfield  {title} {\bibinfo {title} {Nature of the spin
  dynamics and $1/3$ magnetization plateau in azurite},\ }\href
  {https://doi.org/10.1103/PhysRevLett.100.117202} {\bibfield  {journal}
  {\bibinfo  {journal} {Phys. Rev. Lett.}\ }\textbf {\bibinfo {volume} {100}},\
  \bibinfo {pages} {117202} (\bibinfo {year} {2008})}\BibitemShut {NoStop}%
\bibitem [{\citenamefont {Mikeska}\ and\ \citenamefont
  {Luckmann}(2008)}]{PhysRevB.77.054405}%
  \BibitemOpen
  \bibfield  {author} {\bibinfo {author} {\bibfnamefont {H.-J.}\ \bibnamefont
  {Mikeska}}\ and\ \bibinfo {author} {\bibfnamefont {C.}~\bibnamefont
  {Luckmann}},\ }\bibfield  {title} {\bibinfo {title} {Dynamics of a distorted
  diamond chain},\ }\href {https://doi.org/10.1103/PhysRevB.77.054405}
  {\bibfield  {journal} {\bibinfo  {journal} {Phys. Rev. B}\ }\textbf {\bibinfo
  {volume} {77}},\ \bibinfo {pages} {054405} (\bibinfo {year}
  {2008})}\BibitemShut {NoStop}%
\bibitem [{\citenamefont {Gibson}\ \emph {et~al.}(2010)\citenamefont {Gibson},
  \citenamefont {Rule}, \citenamefont {Wolter}, \citenamefont {Hoffmann},
  \citenamefont {Prokhnenko}, \citenamefont {Tennant}, \citenamefont
  {Gerischer}, \citenamefont {Kraken}, \citenamefont {Litterst}, \citenamefont
  {S\"ullow}, \citenamefont {Schreuer}, \citenamefont {Luetkens}, \citenamefont
  {Br\"uhl}, \citenamefont {Wolf},\ and\ \citenamefont
  {Lang}}]{PhysRevB.81.140406}%
  \BibitemOpen
  \bibfield  {author} {\bibinfo {author} {\bibfnamefont {M.~C.~R.}\
  \bibnamefont {Gibson}}, \bibinfo {author} {\bibfnamefont {K.~C.}\
  \bibnamefont {Rule}}, \bibinfo {author} {\bibfnamefont {A.~U.~B.}\
  \bibnamefont {Wolter}}, \bibinfo {author} {\bibfnamefont {J.-U.}\
  \bibnamefont {Hoffmann}}, \bibinfo {author} {\bibfnamefont {O.}~\bibnamefont
  {Prokhnenko}}, \bibinfo {author} {\bibfnamefont {D.~A.}\ \bibnamefont
  {Tennant}}, \bibinfo {author} {\bibfnamefont {S.}~\bibnamefont {Gerischer}},
  \bibinfo {author} {\bibfnamefont {M.}~\bibnamefont {Kraken}}, \bibinfo
  {author} {\bibfnamefont {F.~J.}\ \bibnamefont {Litterst}}, \bibinfo {author}
  {\bibfnamefont {S.}~\bibnamefont {S\"ullow}}, \bibinfo {author}
  {\bibfnamefont {J.}~\bibnamefont {Schreuer}}, \bibinfo {author}
  {\bibfnamefont {H.}~\bibnamefont {Luetkens}}, \bibinfo {author}
  {\bibfnamefont {A.}~\bibnamefont {Br\"uhl}}, \bibinfo {author} {\bibfnamefont
  {B.}~\bibnamefont {Wolf}},\ and\ \bibinfo {author} {\bibfnamefont
  {M.}~\bibnamefont {Lang}},\ }\bibfield  {title} {\bibinfo {title}
  {Magnetoelastic and structural properties of azurite
  \text{Cu$_{3}$}\text{(CO$_{3}$)$_{2}$}\text{(OH)$_{2}$} from neutron
  scattering and muon spin rotation},\ }\href
  {https://doi.org/10.1103/PhysRevB.81.140406} {\bibfield  {journal} {\bibinfo
  {journal} {Phys. Rev. B}\ }\textbf {\bibinfo {volume} {81}},\ \bibinfo
  {pages} {140406} (\bibinfo {year} {2010})}\BibitemShut {NoStop}%
\bibitem [{\citenamefont {Jeschke}\ \emph {et~al.}(2011)\citenamefont
  {Jeschke}, \citenamefont {Opahle}, \citenamefont {Kandpal}, \citenamefont
  {Valent\'{\i}}, \citenamefont {Das}, \citenamefont {Saha-Dasgupta},
  \citenamefont {Janson}, \citenamefont {Rosner}, \citenamefont {Br\"uhl},
  \citenamefont {Wolf}, \citenamefont {Lang}, \citenamefont {Richter},
  \citenamefont {Hu}, \citenamefont {Wang}, \citenamefont {Peters},
  \citenamefont {Pruschke},\ and\ \citenamefont
  {Honecker}}]{PhysRevLett.106.217201}%
  \BibitemOpen
  \bibfield  {author} {\bibinfo {author} {\bibfnamefont {H.}~\bibnamefont
  {Jeschke}}, \bibinfo {author} {\bibfnamefont {I.}~\bibnamefont {Opahle}},
  \bibinfo {author} {\bibfnamefont {H.}~\bibnamefont {Kandpal}}, \bibinfo
  {author} {\bibfnamefont {R.}~\bibnamefont {Valent\'{\i}}}, \bibinfo {author}
  {\bibfnamefont {H.}~\bibnamefont {Das}}, \bibinfo {author} {\bibfnamefont
  {T.}~\bibnamefont {Saha-Dasgupta}}, \bibinfo {author} {\bibfnamefont
  {O.}~\bibnamefont {Janson}}, \bibinfo {author} {\bibfnamefont
  {H.}~\bibnamefont {Rosner}}, \bibinfo {author} {\bibfnamefont
  {A.}~\bibnamefont {Br\"uhl}}, \bibinfo {author} {\bibfnamefont
  {B.}~\bibnamefont {Wolf}}, \bibinfo {author} {\bibfnamefont {M.}~\bibnamefont
  {Lang}}, \bibinfo {author} {\bibfnamefont {J.}~\bibnamefont {Richter}},
  \bibinfo {author} {\bibfnamefont {S.}~\bibnamefont {Hu}}, \bibinfo {author}
  {\bibfnamefont {X.}~\bibnamefont {Wang}}, \bibinfo {author} {\bibfnamefont
  {R.}~\bibnamefont {Peters}}, \bibinfo {author} {\bibfnamefont
  {T.}~\bibnamefont {Pruschke}},\ and\ \bibinfo {author} {\bibfnamefont
  {A.}~\bibnamefont {Honecker}},\ }\bibfield  {title} {\bibinfo {title}
  {Multistep approach to microscopic models for frustrated quantum magnets: The
  case of the natural mineral azurite},\ }\href
  {https://doi.org/10.1103/PhysRevLett.106.217201} {\bibfield  {journal}
  {\bibinfo  {journal} {Phys. Rev. Lett.}\ }\textbf {\bibinfo {volume} {106}},\
  \bibinfo {pages} {217201} (\bibinfo {year} {2011})}\BibitemShut {NoStop}%
\bibitem [{\citenamefont {Kang}\ \emph {et~al.}(2009)\citenamefont {Kang},
  \citenamefont {Lee}, \citenamefont {Kremer},\ and\ \citenamefont
  {Whangbo}}]{Kang_2009}%
  \BibitemOpen
  \bibfield  {author} {\bibinfo {author} {\bibfnamefont {J.}~\bibnamefont
  {Kang}}, \bibinfo {author} {\bibfnamefont {C.}~\bibnamefont {Lee}}, \bibinfo
  {author} {\bibfnamefont {R.~K.}\ \bibnamefont {Kremer}},\ and\ \bibinfo
  {author} {\bibfnamefont {M.-H.}\ \bibnamefont {Whangbo}},\ }\bibfield
  {title} {\bibinfo {title} {Consequences of the intrachain dimer–monomer
  spin frustration and the interchain dimer–monomer spin exchange in the
  diamond-chain compound azurite
  \text{Cu$_{3}$}\text{(CO$_{3}$)$_{2}$}\text{(OH)$_{2}$}},\ }\href
  {https://doi.org/10.1088/0953-8984/21/39/392201} {\bibfield  {journal}
  {\bibinfo  {journal} {J. Condens. Matter Phys.}\ }\textbf {\bibinfo {volume}
  {21}},\ \bibinfo {pages} {392201} (\bibinfo {year} {2009})}\BibitemShut
  {NoStop}%
\bibitem [{\citenamefont {Aimo}\ \emph {et~al.}(2009)\citenamefont {Aimo},
  \citenamefont {Kr\"amer}, \citenamefont {Klanj\ifmmode~\check{s}\else
  \v{s}\fi{}ek}, \citenamefont {Horvati\ifmmode~\acute{c}\else \'{c}\fi{}},
  \citenamefont {Berthier},\ and\ \citenamefont
  {Kikuchi}}]{PhysRevLett.102.127205}%
  \BibitemOpen
  \bibfield  {author} {\bibinfo {author} {\bibfnamefont {F.}~\bibnamefont
  {Aimo}}, \bibinfo {author} {\bibfnamefont {S.}~\bibnamefont {Kr\"amer}},
  \bibinfo {author} {\bibfnamefont {M.}~\bibnamefont
  {Klanj\ifmmode~\check{s}\else \v{s}\fi{}ek}}, \bibinfo {author}
  {\bibfnamefont {M.}~\bibnamefont {Horvati\ifmmode~\acute{c}\else
  \'{c}\fi{}}}, \bibinfo {author} {\bibfnamefont {C.}~\bibnamefont
  {Berthier}},\ and\ \bibinfo {author} {\bibfnamefont {H.}~\bibnamefont
  {Kikuchi}},\ }\bibfield  {title} {\bibinfo {title} {Spin configuration in the
  $1/3$ magnetization plateau of azurite determined by nmr},\ }\href
  {https://doi.org/10.1103/PhysRevLett.102.127205} {\bibfield  {journal}
  {\bibinfo  {journal} {Phys. Rev. Lett.}\ }\textbf {\bibinfo {volume} {102}},\
  \bibinfo {pages} {127205} (\bibinfo {year} {2009})}\BibitemShut {NoStop}%
\bibitem [{\citenamefont {Rule}\ \emph
  {et~al.}(2011{\natexlab{a}})\citenamefont {Rule}, \citenamefont {Tennant},
  \citenamefont {Caux}, \citenamefont {Gibson}, \citenamefont {Telling},
  \citenamefont {Gerischer}, \citenamefont {S\"ullow},\ and\ \citenamefont
  {Lang}}]{PhysRevB.84.184419}%
  \BibitemOpen
  \bibfield  {author} {\bibinfo {author} {\bibfnamefont {K.~C.}\ \bibnamefont
  {Rule}}, \bibinfo {author} {\bibfnamefont {D.~A.}\ \bibnamefont {Tennant}},
  \bibinfo {author} {\bibfnamefont {J.-S.}\ \bibnamefont {Caux}}, \bibinfo
  {author} {\bibfnamefont {M.~C.~R.}\ \bibnamefont {Gibson}}, \bibinfo {author}
  {\bibfnamefont {M.~T.~F.}\ \bibnamefont {Telling}}, \bibinfo {author}
  {\bibfnamefont {S.}~\bibnamefont {Gerischer}}, \bibinfo {author}
  {\bibfnamefont {S.}~\bibnamefont {S\"ullow}},\ and\ \bibinfo {author}
  {\bibfnamefont {M.}~\bibnamefont {Lang}},\ }\bibfield  {title} {\bibinfo
  {title} {Dynamics of azurite
  \text{Cu$_{3}$}\text{(CO$_{3}$)$_{2}$}\text{(OH)$_{2}$} in a magnetic field
  as determined by neutron scattering},\ }\href
  {https://doi.org/10.1103/PhysRevB.84.184419} {\bibfield  {journal} {\bibinfo
  {journal} {Phys. Rev. B}\ }\textbf {\bibinfo {volume} {84}},\ \bibinfo
  {pages} {184419} (\bibinfo {year} {2011}{\natexlab{a}})}\BibitemShut
  {NoStop}%
\bibitem [{\citenamefont {Whangbo}\ \emph {et~al.}(2021)\citenamefont
  {Whangbo}, \citenamefont {Koo},\ and\ \citenamefont
  {Kremer}}]{molecules26030531}%
  \BibitemOpen
  \bibfield  {author} {\bibinfo {author} {\bibfnamefont {M.-H.}\ \bibnamefont
  {Whangbo}}, \bibinfo {author} {\bibfnamefont {H.-J.}\ \bibnamefont {Koo}},\
  and\ \bibinfo {author} {\bibfnamefont {R.~K.}\ \bibnamefont {Kremer}},\
  }\bibfield  {title} {\bibinfo {title} {Spin exchanges between transition
  metal ions governed by the ligand p-orbitals in their magnetic orbitals},\
  }\bibfield  {journal} {\bibinfo  {journal} {Molecules}\ }\textbf {\bibinfo
  {volume} {26}},\ \href {https://doi.org/10.3390/molecules26030531}
  {10.3390/molecules26030531} (\bibinfo {year} {2021})\BibitemShut {NoStop}%
\bibitem [{\citenamefont {Ishii}\ \emph {et~al.}(2000)\citenamefont {Ishii},
  \citenamefont {Tanaka}, \citenamefont {Hori}, \citenamefont {Uekusa},
  \citenamefont {Ohashi}, \citenamefont {Tatani}, \citenamefont {Narumi},\ and\
  \citenamefont {Kindo}}]{ishii2000gapped}%
  \BibitemOpen
  \bibfield  {author} {\bibinfo {author} {\bibfnamefont {M.}~\bibnamefont
  {Ishii}}, \bibinfo {author} {\bibfnamefont {H.}~\bibnamefont {Tanaka}},
  \bibinfo {author} {\bibfnamefont {M.}~\bibnamefont {Hori}}, \bibinfo {author}
  {\bibfnamefont {H.}~\bibnamefont {Uekusa}}, \bibinfo {author} {\bibfnamefont
  {Y.}~\bibnamefont {Ohashi}}, \bibinfo {author} {\bibfnamefont
  {K.}~\bibnamefont {Tatani}}, \bibinfo {author} {\bibfnamefont
  {Y.}~\bibnamefont {Narumi}},\ and\ \bibinfo {author} {\bibfnamefont
  {K.}~\bibnamefont {Kindo}},\ }\bibfield  {title} {\bibinfo {title} {Gapped
  ground state in the spin-1/2 trimer chain system
  \text{Cu$_{3}$}\text{Cl$_{6}$}\text{(H$_{2}$O)$_{2}$}·\text{2H}
  \text{8}\text{C$_{4}$}\text{S}\text{O$_{2}$}},\ }\href@noop {} {\bibfield
  {journal} {\bibinfo  {journal} {J. Phys. Soc. Jpn.}\ }\textbf {\bibinfo
  {volume} {69}},\ \bibinfo {pages} {340} (\bibinfo {year} {2000})}\BibitemShut
  {NoStop}%
\bibitem [{\citenamefont {Hida}\ and\ \citenamefont
  {Takano}(2011)}]{doi:10.1143/JPSJ.80.104710}%
  \BibitemOpen
  \bibfield  {author} {\bibinfo {author} {\bibfnamefont {K.}~\bibnamefont
  {Hida}}\ and\ \bibinfo {author} {\bibfnamefont {K.}~\bibnamefont {Takano}},\
  }\bibfield  {title} {\bibinfo {title} {Effects of single-site anisotropy on
  mixed diamond chains with spins 1 and 1/2},\ }\href
  {https://doi.org/10.1143/JPSJ.80.104710} {\bibfield  {journal} {\bibinfo
  {journal} {J. Phys. Soc. Jpn.}\ }\textbf {\bibinfo {volume} {80}},\ \bibinfo
  {pages} {104710} (\bibinfo {year} {2011})}\BibitemShut {NoStop}%
\bibitem [{\citenamefont {Fujihala}\ \emph {et~al.}(2015)\citenamefont
  {Fujihala}, \citenamefont {Koorikawa}, \citenamefont {Mitsuda}, \citenamefont
  {Hagihala}, \citenamefont {Morodomi}, \citenamefont {Kawae}, \citenamefont
  {Matsuo},\ and\ \citenamefont {Kindo}}]{doi:10.7566/JPSJ.84.073702}%
  \BibitemOpen
  \bibfield  {author} {\bibinfo {author} {\bibfnamefont {M.}~\bibnamefont
  {Fujihala}}, \bibinfo {author} {\bibfnamefont {H.}~\bibnamefont {Koorikawa}},
  \bibinfo {author} {\bibfnamefont {S.}~\bibnamefont {Mitsuda}}, \bibinfo
  {author} {\bibfnamefont {M.}~\bibnamefont {Hagihala}}, \bibinfo {author}
  {\bibfnamefont {H.}~\bibnamefont {Morodomi}}, \bibinfo {author}
  {\bibfnamefont {T.}~\bibnamefont {Kawae}}, \bibinfo {author} {\bibfnamefont
  {A.}~\bibnamefont {Matsuo}},\ and\ \bibinfo {author} {\bibfnamefont
  {K.}~\bibnamefont {Kindo}},\ }\bibfield  {title} {\bibinfo {title}
  {Spin-liquid ground state in the spin 1/2 distorted diamond chain compound
  \text{K$_{3}$}\text{Cu$_{3}$}\text{Al}\text{O$_{2}$}\text{(SO$_{4}$)$_{4}$}},\
  }\href {https://doi.org/10.7566/JPSJ.84.073702} {\bibfield  {journal}
  {\bibinfo  {journal} {J. Phys. Soc. Jpn.}\ }\textbf {\bibinfo {volume}
  {84}},\ \bibinfo {pages} {073702} (\bibinfo {year} {2015})}\BibitemShut
  {NoStop}%
\bibitem [{\citenamefont {Perdew}\ \emph {et~al.}(1996)\citenamefont {Perdew},
  \citenamefont {Burke},\ and\ \citenamefont
  {Ernzerhof}}]{PhysRevLett.77.3865}%
  \BibitemOpen
  \bibfield  {author} {\bibinfo {author} {\bibfnamefont {J.~P.}\ \bibnamefont
  {Perdew}}, \bibinfo {author} {\bibfnamefont {K.}~\bibnamefont {Burke}},\ and\
  \bibinfo {author} {\bibfnamefont {M.}~\bibnamefont {Ernzerhof}},\ }\bibfield
  {title} {\bibinfo {title} {Generalized gradient approximation made simple},\
  }\href {https://doi.org/10.1103/PhysRevLett.77.3865} {\bibfield  {journal}
  {\bibinfo  {journal} {Phys. Rev. Lett.}\ }\textbf {\bibinfo {volume} {77}},\
  \bibinfo {pages} {3865} (\bibinfo {year} {1996})}\BibitemShut {NoStop}%
\bibitem [{\citenamefont {Dudarev}\ \emph {et~al.}(1998)\citenamefont
  {Dudarev}, \citenamefont {Botton}, \citenamefont {Savrasov}, \citenamefont
  {Humphreys},\ and\ \citenamefont {Sutton}}]{PhysRevB.57.1505}%
  \BibitemOpen
  \bibfield  {author} {\bibinfo {author} {\bibfnamefont {S.~L.}\ \bibnamefont
  {Dudarev}}, \bibinfo {author} {\bibfnamefont {G.~A.}\ \bibnamefont {Botton}},
  \bibinfo {author} {\bibfnamefont {S.~Y.}\ \bibnamefont {Savrasov}}, \bibinfo
  {author} {\bibfnamefont {C.~J.}\ \bibnamefont {Humphreys}},\ and\ \bibinfo
  {author} {\bibfnamefont {A.~P.}\ \bibnamefont {Sutton}},\ }\bibfield  {title}
  {\bibinfo {title} {Electron-energy-loss spectra and the structural stability
  of nickel oxide: An lsda+u study},\ }\href
  {https://doi.org/10.1103/PhysRevB.57.1505} {\bibfield  {journal} {\bibinfo
  {journal} {Phys. Rev. B}\ }\textbf {\bibinfo {volume} {57}},\ \bibinfo
  {pages} {1505} (\bibinfo {year} {1998})}\BibitemShut {NoStop}%
\bibitem [{\citenamefont {Petersen}\ \emph {et~al.}(2006)\citenamefont
  {Petersen}, \citenamefont {Hafner},\ and\ \citenamefont
  {Marsman}}]{Petersen2006}%
  \BibitemOpen
  \bibfield  {author} {\bibinfo {author} {\bibfnamefont {M.}~\bibnamefont
  {Petersen}}, \bibinfo {author} {\bibfnamefont {J.}~\bibnamefont {Hafner}},\
  and\ \bibinfo {author} {\bibfnamefont {M.}~\bibnamefont {Marsman}},\
  }\bibfield  {title} {\bibinfo {title} {Structural, electronic and magnetic
  properties of gd investigated by dft+u methods: bulk, clean and h-covered
  (0001) surfaces},\ }\href@noop {} {\bibfield  {journal} {\bibinfo  {journal}
  {J. Condens. Matter Phys.}\ }\textbf {\bibinfo {volume} {18}},\ \bibinfo
  {pages} {7021} (\bibinfo {year} {2006})}\BibitemShut {NoStop}%
\bibitem [{\citenamefont {Kresse}\ and\ \citenamefont
  {Joubert}(1999)}]{PhysRevB.59.1758}%
  \BibitemOpen
  \bibfield  {author} {\bibinfo {author} {\bibfnamefont {G.}~\bibnamefont
  {Kresse}}\ and\ \bibinfo {author} {\bibfnamefont {D.}~\bibnamefont
  {Joubert}},\ }\bibfield  {title} {\bibinfo {title} {From ultrasoft
  pseudopotentials to the projector augmented-wave method},\ }\href
  {https://doi.org/10.1103/PhysRevB.59.1758} {\bibfield  {journal} {\bibinfo
  {journal} {Phys. Rev. B}\ }\textbf {\bibinfo {volume} {59}},\ \bibinfo
  {pages} {1758} (\bibinfo {year} {1999})}\BibitemShut {NoStop}%
\bibitem [{\citenamefont {Kresse}\ and\ \citenamefont
  {Hafner}(1993)}]{PhysRevB.47.558}%
  \BibitemOpen
  \bibfield  {author} {\bibinfo {author} {\bibfnamefont {G.}~\bibnamefont
  {Kresse}}\ and\ \bibinfo {author} {\bibfnamefont {J.}~\bibnamefont
  {Hafner}},\ }\bibfield  {title} {\bibinfo {title} {Ab initio molecular
  dynamics for liquid metals},\ }\href
  {https://doi.org/10.1103/PhysRevB.47.558} {\bibfield  {journal} {\bibinfo
  {journal} {Phys. Rev. B}\ }\textbf {\bibinfo {volume} {47}},\ \bibinfo
  {pages} {558} (\bibinfo {year} {1993})}\BibitemShut {NoStop}%
\bibitem [{\citenamefont {Kresse}\ and\ \citenamefont
  {Hafner}(1994)}]{PhysRevB.49.14251}%
  \BibitemOpen
  \bibfield  {author} {\bibinfo {author} {\bibfnamefont {G.}~\bibnamefont
  {Kresse}}\ and\ \bibinfo {author} {\bibfnamefont {J.}~\bibnamefont
  {Hafner}},\ }\bibfield  {title} {\bibinfo {title} {Ab initio
  molecular-dynamics simulation of the liquid-metal--amorphous-semiconductor
  transition in germanium},\ }\href {https://doi.org/10.1103/PhysRevB.49.14251}
  {\bibfield  {journal} {\bibinfo  {journal} {Phys. Rev. B}\ }\textbf {\bibinfo
  {volume} {49}},\ \bibinfo {pages} {14251} (\bibinfo {year}
  {1994})}\BibitemShut {NoStop}%
\bibitem [{\citenamefont {Liechtenstein}\ \emph {et~al.}(1987)\citenamefont
  {Liechtenstein}, \citenamefont {Katsnelson}, \citenamefont {Antropov},\ and\
  \citenamefont {Gubanov}}]{LIECHTENSTEIN198765}%
  \BibitemOpen
  \bibfield  {author} {\bibinfo {author} {\bibfnamefont {A.}~\bibnamefont
  {Liechtenstein}}, \bibinfo {author} {\bibfnamefont {M.}~\bibnamefont
  {Katsnelson}}, \bibinfo {author} {\bibfnamefont {V.}~\bibnamefont
  {Antropov}},\ and\ \bibinfo {author} {\bibfnamefont {V.}~\bibnamefont
  {Gubanov}},\ }\bibfield  {title} {\bibinfo {title} {Local spin density
  functional approach to the theory of exchange interactions in ferromagnetic
  metals and alloys},\ }\href
  {https://doi.org/https://doi.org/10.1016/0304-8853(87)90721-9} {\bibfield
  {journal} {\bibinfo  {journal} {J. Magn. Magn. Mater.}\ }\textbf {\bibinfo
  {volume} {67}},\ \bibinfo {pages} {65} (\bibinfo {year} {1987})}\BibitemShut
  {NoStop}%
\bibitem [{\citenamefont {Ozaki}(2003)}]{PhysRevB.67.155108}%
  \BibitemOpen
  \bibfield  {author} {\bibinfo {author} {\bibfnamefont {T.}~\bibnamefont
  {Ozaki}},\ }\bibfield  {title} {\bibinfo {title} {Variationally optimized
  atomic orbitals for large-scale electronic structures},\ }\href
  {https://doi.org/10.1103/PhysRevB.67.155108} {\bibfield  {journal} {\bibinfo
  {journal} {Phys. Rev. B}\ }\textbf {\bibinfo {volume} {67}},\ \bibinfo
  {pages} {155108} (\bibinfo {year} {2003})}\BibitemShut {NoStop}%
\bibitem [{\citenamefont {Yoon}\ \emph {et~al.}(2020)\citenamefont {Yoon},
  \citenamefont {Kim}, \citenamefont {Sim},\ and\ \citenamefont
  {Han}}]{YOON2020106927}%
  \BibitemOpen
  \bibfield  {author} {\bibinfo {author} {\bibfnamefont {H.}~\bibnamefont
  {Yoon}}, \bibinfo {author} {\bibfnamefont {T.~J.}\ \bibnamefont {Kim}},
  \bibinfo {author} {\bibfnamefont {J.-H.}\ \bibnamefont {Sim}},\ and\ \bibinfo
  {author} {\bibfnamefont {M.~J.}\ \bibnamefont {Han}},\ }\bibfield  {title}
  {\bibinfo {title} {Jx: An open-source software for calculating magnetic
  interactions based on magnetic force theory},\ }\href
  {https://doi.org/https://doi.org/10.1016/j.cpc.2019.106927} {\bibfield
  {journal} {\bibinfo  {journal} {Comput. Phys. Commun.}\ }\textbf {\bibinfo
  {volume} {247}},\ \bibinfo {pages} {106927} (\bibinfo {year}
  {2020})}\BibitemShut {NoStop}%
\bibitem [{\citenamefont {Rodriguez-Carvajal}(1990)}]{rodriguez1990fullprof}%
  \BibitemOpen
  \bibfield  {author} {\bibinfo {author} {\bibfnamefont {J.}~\bibnamefont
  {Rodriguez-Carvajal}},\ }\bibfield  {title} {\bibinfo {title} {Fullprof: a
  program for rietveld refinement and pattern matching analysis},\ }in\
  \href@noop {} {\emph {\bibinfo {booktitle} {satellite meeting on powder
  diffraction of the XV congress of the IUCr}}},\ Vol.\ \bibinfo {volume}
  {127}\ (\bibinfo {organization} {Toulouse, France:[sn]},\ \bibinfo {year}
  {1990})\BibitemShut {NoStop}%
\bibitem [{sm()}]{sm}%
  \BibitemOpen
  \href@noop {} {}\bibinfo {note} {L. A. Glinskaya, R. F. Klevtsova, V. G. Kim,
  and P. V.Klevtsov, The synthesis and the crystal structure of the double
  molybdate \text{K$_{2}$}\text{Cu$_{3}$}\text{(MoO$_{4}$)$_{4}$}, Soviet
  Physics Doklady, \textbf{25}, 794, (1980)}\BibitemShut {NoStop}%
\bibitem [{\citenamefont {Rule}\ \emph
  {et~al.}(2011{\natexlab{b}})\citenamefont {Rule}, \citenamefont {Reehuis},
  \citenamefont {Gibson}, \citenamefont {Ouladdiaf}, \citenamefont {Gutmann},
  \citenamefont {Hoffmann}, \citenamefont {Gerischer}, \citenamefont {Tennant},
  \citenamefont {S\"ullow},\ and\ \citenamefont {Lang}}]{PhysRevB.83.104401}%
  \BibitemOpen
  \bibfield  {author} {\bibinfo {author} {\bibfnamefont {K.~C.}\ \bibnamefont
  {Rule}}, \bibinfo {author} {\bibfnamefont {M.}~\bibnamefont {Reehuis}},
  \bibinfo {author} {\bibfnamefont {M.~C.~R.}\ \bibnamefont {Gibson}}, \bibinfo
  {author} {\bibfnamefont {B.}~\bibnamefont {Ouladdiaf}}, \bibinfo {author}
  {\bibfnamefont {M.~J.}\ \bibnamefont {Gutmann}}, \bibinfo {author}
  {\bibfnamefont {J.-U.}\ \bibnamefont {Hoffmann}}, \bibinfo {author}
  {\bibfnamefont {S.}~\bibnamefont {Gerischer}}, \bibinfo {author}
  {\bibfnamefont {D.~A.}\ \bibnamefont {Tennant}}, \bibinfo {author}
  {\bibfnamefont {S.}~\bibnamefont {S\"ullow}},\ and\ \bibinfo {author}
  {\bibfnamefont {M.}~\bibnamefont {Lang}},\ }\bibfield  {title} {\bibinfo
  {title} {Magnetic and crystal structure of azurite
  \text{Cu$_{3}$}\text{(CO$_{3}$)$_{2}$}\text{(OH)$_{2}$} as determined by
  neutron diffraction},\ }\href {https://doi.org/10.1103/PhysRevB.83.104401}
  {\bibfield  {journal} {\bibinfo  {journal} {Phys. Rev. B}\ }\textbf {\bibinfo
  {volume} {83}},\ \bibinfo {pages} {104401} (\bibinfo {year}
  {2011}{\natexlab{b}})}\BibitemShut {NoStop}%
\bibitem [{\citenamefont {Bain}\ and\ \citenamefont {Berry}(2008)}]{Bain2008}%
  \BibitemOpen
  \bibfield  {author} {\bibinfo {author} {\bibfnamefont {G.~A.}\ \bibnamefont
  {Bain}}\ and\ \bibinfo {author} {\bibfnamefont {J.~F.}\ \bibnamefont
  {Berry}},\ }\bibfield  {title} {\bibinfo {title} {Diamagnetic corrections and
  pascal's constants},\ }\href {https://doi.org/10.1021/ed085p532} {\bibfield
  {journal} {\bibinfo  {journal} {J. Chem. Educ.}\ }\textbf {\bibinfo {volume}
  {85}},\ \bibinfo {pages} {532} (\bibinfo {year} {2008})}\BibitemShut
  {NoStop}%
\bibitem [{\citenamefont {Narsinga~Rao}\ \emph {et~al.}(2015)\citenamefont
  {Narsinga~Rao}, \citenamefont {Singh}, \citenamefont {Sankar}, \citenamefont
  {Muthuselvam}, \citenamefont {Guo},\ and\ \citenamefont
  {Chou}}]{PhysRevB.91.014423}%
  \BibitemOpen
  \bibfield  {author} {\bibinfo {author} {\bibfnamefont {G.}~\bibnamefont
  {Narsinga~Rao}}, \bibinfo {author} {\bibfnamefont {V.~N.}\ \bibnamefont
  {Singh}}, \bibinfo {author} {\bibfnamefont {R.}~\bibnamefont {Sankar}},
  \bibinfo {author} {\bibfnamefont {I.~P.}\ \bibnamefont {Muthuselvam}},
  \bibinfo {author} {\bibfnamefont {G.-Y.}\ \bibnamefont {Guo}},\ and\ \bibinfo
  {author} {\bibfnamefont {F.~C.}\ \bibnamefont {Chou}},\ }\bibfield  {title}
  {\bibinfo {title} {Antiferromagnetism of
  \text{Ni$_{2}$}\text{Nb}\text{B}\text{O$_{6}$} with $s=1$ dimer
  quasi-one-dimensional armchair chains},\ }\href
  {https://doi.org/10.1103/PhysRevB.91.014423} {\bibfield  {journal} {\bibinfo
  {journal} {Phys. Rev. B}\ }\textbf {\bibinfo {volume} {91}},\ \bibinfo
  {pages} {014423} (\bibinfo {year} {2015})}\BibitemShut {NoStop}%
\bibitem [{\citenamefont {Sakai}\ \emph {et~al.}(2022)\citenamefont {Sakai},
  \citenamefont {Okamoto}, \citenamefont {Nakano},\ and\ \citenamefont
  {Furuchi}}]{10.1063/9.0000255}%
  \BibitemOpen
  \bibfield  {author} {\bibinfo {author} {\bibfnamefont {T.}~\bibnamefont
  {Sakai}}, \bibinfo {author} {\bibfnamefont {K.}~\bibnamefont {Okamoto}},
  \bibinfo {author} {\bibfnamefont {H.}~\bibnamefont {Nakano}},\ and\ \bibinfo
  {author} {\bibfnamefont {R.}~\bibnamefont {Furuchi}},\ }\bibfield  {title}
  {\bibinfo {title} {{Magnetization plateau of the distorted diamond spin chain
  with anisotropic ferromagnetic interaction}},\ }\href
  {https://doi.org/10.1063/9.0000255} {\bibfield  {journal} {\bibinfo
  {journal} {AIP Advances}\ }\textbf {\bibinfo {volume} {12}},\ \bibinfo
  {pages} {035030} (\bibinfo {year} {2022})}\BibitemShut {NoStop}%
\bibitem [{\citenamefont {Sachdev}(2000)}]{Subir}%
  \BibitemOpen
  \bibfield  {author} {\bibinfo {author} {\bibfnamefont {S.}~\bibnamefont
  {Sachdev}},\ }\bibfield  {title} {\bibinfo {title} {Quantum criticality:
  Competing ground states in low dimensions},\ }\href
  {https://doi.org/10.1126/science.288.5465.475} {\bibfield  {journal}
  {\bibinfo  {journal} {Science}\ }\textbf {\bibinfo {volume} {288}},\ \bibinfo
  {pages} {475} (\bibinfo {year} {2000})}\BibitemShut {NoStop}%
\bibitem [{\citenamefont {Lake}\ \emph {et~al.}(2005)\citenamefont {Lake},
  \citenamefont {Tennant}, \citenamefont {Frost},\ and\ \citenamefont
  {Nagler}}]{Lake2005}%
  \BibitemOpen
  \bibfield  {author} {\bibinfo {author} {\bibfnamefont {B.}~\bibnamefont
  {Lake}}, \bibinfo {author} {\bibfnamefont {D.~A.}\ \bibnamefont {Tennant}},
  \bibinfo {author} {\bibfnamefont {C.~D.}\ \bibnamefont {Frost}},\ and\
  \bibinfo {author} {\bibfnamefont {S.~E.}\ \bibnamefont {Nagler}},\ }\bibfield
   {title} {\bibinfo {title} {Quantum criticality and universal scaling of a
  quantum antiferromagnet},\ }\href {https://doi.org/10.1038/nmat1327}
  {\bibfield  {journal} {\bibinfo  {journal} {Nat. Mater.}\ }\textbf {\bibinfo
  {volume} {4}},\ \bibinfo {pages} {329} (\bibinfo {year} {2005})}\BibitemShut
  {NoStop}%
\bibitem [{\citenamefont {Amaya}\ \emph {et~al.}(2017)\citenamefont {Amaya},
  \citenamefont {Ono}, \citenamefont {Oku}, \citenamefont {Yamaguchi},
  \citenamefont {Matsuo}, \citenamefont {Kindo}, \citenamefont {Nojiri},
  \citenamefont {Palacio}, \citenamefont {Campo},\ and\ \citenamefont
  {Hosokoshi}}]{doi:10.7566/JPSJ.86.074706}%
  \BibitemOpen
  \bibfield  {author} {\bibinfo {author} {\bibfnamefont {N.}~\bibnamefont
  {Amaya}}, \bibinfo {author} {\bibfnamefont {T.}~\bibnamefont {Ono}}, \bibinfo
  {author} {\bibfnamefont {Y.}~\bibnamefont {Oku}}, \bibinfo {author}
  {\bibfnamefont {H.}~\bibnamefont {Yamaguchi}}, \bibinfo {author}
  {\bibfnamefont {A.}~\bibnamefont {Matsuo}}, \bibinfo {author} {\bibfnamefont
  {K.}~\bibnamefont {Kindo}}, \bibinfo {author} {\bibfnamefont
  {H.}~\bibnamefont {Nojiri}}, \bibinfo {author} {\bibfnamefont
  {F.}~\bibnamefont {Palacio}}, \bibinfo {author} {\bibfnamefont
  {J.}~\bibnamefont {Campo}},\ and\ \bibinfo {author} {\bibfnamefont
  {Y.}~\bibnamefont {Hosokoshi}},\ }\bibfield  {title} {\bibinfo {title}
  {Spin-1/2 quantum antiferromagnet on a three-dimensional honeycomb lattice
  formed by a new organic biradical \text{F$_{4}$}{BIPBNN}},\ }\href
  {https://doi.org/10.7566/JPSJ.86.074706} {\bibfield  {journal} {\bibinfo
  {journal} {J. Phys. Soc. Jpn.}\ }\textbf {\bibinfo {volume} {86}},\ \bibinfo
  {pages} {074706} (\bibinfo {year} {2017})}\BibitemShut {NoStop}%
\bibitem [{\citenamefont {White}\ \emph {et~al.}(1983)\citenamefont {White},
  \citenamefont {White},\ and\ \citenamefont {Bayne}}]{white1983quantum}%
  \BibitemOpen
  \bibfield  {author} {\bibinfo {author} {\bibfnamefont {R.~M.}\ \bibnamefont
  {White}}, \bibinfo {author} {\bibfnamefont {R.~M.}\ \bibnamefont {White}},\
  and\ \bibinfo {author} {\bibfnamefont {B.}~\bibnamefont {Bayne}},\
  }\href@noop {} {\emph {\bibinfo {title} {Quantum theory of magnetism}}},\
  Vol.~\bibinfo {volume} {1}\ (\bibinfo  {publisher} {Springer},\ \bibinfo
  {year} {1983})\BibitemShut {NoStop}%
\bibitem [{\citenamefont {Bera}\ \emph {et~al.}(2020)\citenamefont {Bera},
  \citenamefont {Wu}, \citenamefont {Yang}, \citenamefont {Bewley},
  \citenamefont {Boehm}, \citenamefont {Xu}, \citenamefont {Bartkowiak},
  \citenamefont {Prokhnenko}, \citenamefont {Klemke}, \citenamefont {Islam},
  \citenamefont {Law}, \citenamefont {Wang},\ and\ \citenamefont
  {Lake}}]{Bera2020}%
  \BibitemOpen
  \bibfield  {author} {\bibinfo {author} {\bibfnamefont {A.~K.}\ \bibnamefont
  {Bera}}, \bibinfo {author} {\bibfnamefont {J.}~\bibnamefont {Wu}}, \bibinfo
  {author} {\bibfnamefont {W.}~\bibnamefont {Yang}}, \bibinfo {author}
  {\bibfnamefont {R.}~\bibnamefont {Bewley}}, \bibinfo {author} {\bibfnamefont
  {M.}~\bibnamefont {Boehm}}, \bibinfo {author} {\bibfnamefont
  {J.}~\bibnamefont {Xu}}, \bibinfo {author} {\bibfnamefont {M.}~\bibnamefont
  {Bartkowiak}}, \bibinfo {author} {\bibfnamefont {O.}~\bibnamefont
  {Prokhnenko}}, \bibinfo {author} {\bibfnamefont {B.}~\bibnamefont {Klemke}},
  \bibinfo {author} {\bibfnamefont {A.~T. M.~N.}\ \bibnamefont {Islam}},
  \bibinfo {author} {\bibfnamefont {J.~M.}\ \bibnamefont {Law}}, \bibinfo
  {author} {\bibfnamefont {Z.}~\bibnamefont {Wang}},\ and\ \bibinfo {author}
  {\bibfnamefont {B.}~\bibnamefont {Lake}},\ }\bibfield  {title} {\bibinfo
  {title} {Dispersions of many-body bethe strings},\ }\href
  {https://doi.org/10.1038/s41567-020-0835-7} {\bibfield  {journal} {\bibinfo
  {journal} {Nat. Phys.}\ }\textbf {\bibinfo {volume} {16}},\ \bibinfo {pages}
  {625} (\bibinfo {year} {2020})}\BibitemShut {NoStop}%
\bibitem [{\citenamefont {Wu}\ \emph {et~al.}(2019)\citenamefont {Wu},
  \citenamefont {Nikitin}, \citenamefont {Wang}, \citenamefont {Zhu},
  \citenamefont {Batista}, \citenamefont {Tsvelik}, \citenamefont {Samarakoon},
  \citenamefont {Tennant}, \citenamefont {Brando}, \citenamefont {Vasylechko},
  \citenamefont {Frontzek}, \citenamefont {Savici}, \citenamefont {Sala},
  \citenamefont {Ehlers}, \citenamefont {Christianson}, \citenamefont
  {Lumsden},\ and\ \citenamefont {Podlesnyak}}]{Wu2019}%
  \BibitemOpen
  \bibfield  {author} {\bibinfo {author} {\bibfnamefont {L.~S.}\ \bibnamefont
  {Wu}}, \bibinfo {author} {\bibfnamefont {S.~E.}\ \bibnamefont {Nikitin}},
  \bibinfo {author} {\bibfnamefont {Z.}~\bibnamefont {Wang}}, \bibinfo {author}
  {\bibfnamefont {W.}~\bibnamefont {Zhu}}, \bibinfo {author} {\bibfnamefont
  {C.~D.}\ \bibnamefont {Batista}}, \bibinfo {author} {\bibfnamefont {A.~M.}\
  \bibnamefont {Tsvelik}}, \bibinfo {author} {\bibfnamefont {A.~M.}\
  \bibnamefont {Samarakoon}}, \bibinfo {author} {\bibfnamefont {D.~A.}\
  \bibnamefont {Tennant}}, \bibinfo {author} {\bibfnamefont {M.}~\bibnamefont
  {Brando}}, \bibinfo {author} {\bibfnamefont {L.}~\bibnamefont {Vasylechko}},
  \bibinfo {author} {\bibfnamefont {M.}~\bibnamefont {Frontzek}}, \bibinfo
  {author} {\bibfnamefont {A.~T.}\ \bibnamefont {Savici}}, \bibinfo {author}
  {\bibfnamefont {G.}~\bibnamefont {Sala}}, \bibinfo {author} {\bibfnamefont
  {G.}~\bibnamefont {Ehlers}}, \bibinfo {author} {\bibfnamefont {A.~D.}\
  \bibnamefont {Christianson}}, \bibinfo {author} {\bibfnamefont {M.~D.}\
  \bibnamefont {Lumsden}},\ and\ \bibinfo {author} {\bibfnamefont
  {A.}~\bibnamefont {Podlesnyak}},\ }\bibfield  {title} {\bibinfo {title}
  {Tomonaga--luttinger liquid behavior and spinon confinement in ybalo3},\
  }\href {https://doi.org/10.1038/s41467-019-08485-7} {\bibfield  {journal}
  {\bibinfo  {journal} {Nat. Commun.}\ }\textbf {\bibinfo {volume} {10}},\
  \bibinfo {pages} {698} (\bibinfo {year} {2019})}\BibitemShut {NoStop}%
\bibitem [{\citenamefont {Mokhtari}\ \emph {et~al.}(2024)\citenamefont
  {Mokhtari}, \citenamefont {Galeski}, \citenamefont {Stockert}, \citenamefont
  {Nikitin}, \citenamefont {Wawrzynczak}, \citenamefont {Kuechler},
  \citenamefont {Brando}, \citenamefont {Vasylechko}, \citenamefont {Starykh},\
  and\ \citenamefont {Hassinger}}]{mokht}%
  \BibitemOpen
  \bibfield  {author} {\bibinfo {author} {\bibfnamefont {P.}~\bibnamefont
  {Mokhtari}}, \bibinfo {author} {\bibfnamefont {S.}~\bibnamefont {Galeski}},
  \bibinfo {author} {\bibfnamefont {U.}~\bibnamefont {Stockert}}, \bibinfo
  {author} {\bibfnamefont {S.~E.}\ \bibnamefont {Nikitin}}, \bibinfo {author}
  {\bibfnamefont {R.}~\bibnamefont {Wawrzynczak}}, \bibinfo {author}
  {\bibfnamefont {R.}~\bibnamefont {Kuechler}}, \bibinfo {author}
  {\bibfnamefont {M.}~\bibnamefont {Brando}}, \bibinfo {author} {\bibfnamefont
  {L.}~\bibnamefont {Vasylechko}}, \bibinfo {author} {\bibfnamefont {O.~A.}\
  \bibnamefont {Starykh}},\ and\ \bibinfo {author} {\bibfnamefont
  {E.}~\bibnamefont {Hassinger}},\ }\href {https://arxiv.org/abs/2412.21144}
  {\bibinfo {title} {1/5 and 1/3 magnetization plateaux in the spin 1/2 chain
  system ybalo3}} (\bibinfo {year} {2024}),\ \Eprint
  {https://arxiv.org/abs/2412.21144} {arXiv:2412.21144 [cond-mat.str-el]}
  \BibitemShut {NoStop}%
\bibitem [{\citenamefont {Kittel}\ and\ \citenamefont
  {McEuen}(2018)}]{kittel2018introduction}%
  \BibitemOpen
  \bibfield  {author} {\bibinfo {author} {\bibfnamefont {C.}~\bibnamefont
  {Kittel}}\ and\ \bibinfo {author} {\bibfnamefont {P.}~\bibnamefont
  {McEuen}},\ }\href@noop {} {\emph {\bibinfo {title} {Introduction to solid
  state physics}}}\ (\bibinfo  {publisher} {John Wiley \& Sons},\ \bibinfo
  {year} {2018})\BibitemShut {NoStop}%
\bibitem [{\citenamefont {Khatua}\ \emph {et~al.}(2021)\citenamefont {Khatua},
  \citenamefont {Arh}, \citenamefont {Mishra}, \citenamefont {Luetkens},
  \citenamefont {Zorko}, \citenamefont {Sana}, \citenamefont {Rao},
  \citenamefont {Nanda},\ and\ \citenamefont {Khuntia}}]{Khatua2021}%
  \BibitemOpen
  \bibfield  {author} {\bibinfo {author} {\bibfnamefont {J.}~\bibnamefont
  {Khatua}}, \bibinfo {author} {\bibfnamefont {T.}~\bibnamefont {Arh}},
  \bibinfo {author} {\bibfnamefont {S.~B.}\ \bibnamefont {Mishra}}, \bibinfo
  {author} {\bibfnamefont {H.}~\bibnamefont {Luetkens}}, \bibinfo {author}
  {\bibfnamefont {A.}~\bibnamefont {Zorko}}, \bibinfo {author} {\bibfnamefont
  {B.}~\bibnamefont {Sana}}, \bibinfo {author} {\bibfnamefont {M.~S.~R.}\
  \bibnamefont {Rao}}, \bibinfo {author} {\bibfnamefont {B.~R.~K.}\
  \bibnamefont {Nanda}},\ and\ \bibinfo {author} {\bibfnamefont
  {P.}~\bibnamefont {Khuntia}},\ }\bibfield  {title} {\bibinfo {title}
  {Development of short and long-range magnetic order in the double perovskite
  based frustrated triangular lattice antiferromagnet
  \text{Ba$_{2}$}\text{MnTe}\text{O$_{6}$}},\ }\href
  {https://doi.org/10.1038/s41598-021-84876-5} {\bibfield  {journal} {\bibinfo
  {journal} {Sci. Rep.}\ }\textbf {\bibinfo {volume} {11}},\ \bibinfo {pages}
  {6959} (\bibinfo {year} {2021})}\BibitemShut {NoStop}%
\bibitem [{\citenamefont {Svoboda}\ \emph {et~al.}(2001)\citenamefont
  {Svoboda}, \citenamefont {Javorsk\'y}, \citenamefont
  {Divi\ifmmode~\check{s}\else \v{s}\fi{}}, \citenamefont {Sechovsk\'y},
  \citenamefont {Honda}, \citenamefont {Oomi},\ and\ \citenamefont
  {Menovsky}}]{PhysRevB.63.212408}%
  \BibitemOpen
  \bibfield  {author} {\bibinfo {author} {\bibfnamefont {P.}~\bibnamefont
  {Svoboda}}, \bibinfo {author} {\bibfnamefont {P.}~\bibnamefont {Javorsk\'y}},
  \bibinfo {author} {\bibfnamefont {M.}~\bibnamefont
  {Divi\ifmmode~\check{s}\else \v{s}\fi{}}}, \bibinfo {author} {\bibfnamefont
  {V.}~\bibnamefont {Sechovsk\'y}}, \bibinfo {author} {\bibfnamefont
  {F.}~\bibnamefont {Honda}}, \bibinfo {author} {\bibfnamefont
  {G.}~\bibnamefont {Oomi}},\ and\ \bibinfo {author} {\bibfnamefont {A.~A.}\
  \bibnamefont {Menovsky}},\ }\bibfield  {title} {\bibinfo {title} {Importance
  of anharmonic terms in the analysis of the specific heat of
  \text{U}\text{Ni$_{2}$}\text{Si$_{2}$}},\ }\href
  {https://doi.org/10.1103/PhysRevB.63.212408} {\bibfield  {journal} {\bibinfo
  {journal} {Phys. Rev. B}\ }\textbf {\bibinfo {volume} {63}},\ \bibinfo
  {pages} {212408} (\bibinfo {year} {2001})}\BibitemShut {NoStop}%
\bibitem [{\citenamefont {Khatua}\ \emph {et~al.}(2024)\citenamefont {Khatua},
  \citenamefont {Gomil\ifmmode~\check{s}\else \v{s}\fi{}ek}, \citenamefont
  {Choi},\ and\ \citenamefont {Khuntia}}]{PhysRevB.110.184402}%
  \BibitemOpen
  \bibfield  {author} {\bibinfo {author} {\bibfnamefont {J.}~\bibnamefont
  {Khatua}}, \bibinfo {author} {\bibfnamefont {M.}~\bibnamefont
  {Gomil\ifmmode~\check{s}\else \v{s}\fi{}ek}}, \bibinfo {author}
  {\bibfnamefont {K.-Y.}\ \bibnamefont {Choi}},\ and\ \bibinfo {author}
  {\bibfnamefont {P.}~\bibnamefont {Khuntia}},\ }\bibfield  {title} {\bibinfo
  {title} {Magnetism and field-induced effects in the $s=\frac{5}{2}$ honeycomb
  lattice antiferromagnet \text{Fe}\text{P$_{3}$}\text{Si}\text{O$_{11}$}},\
  }\href {https://doi.org/10.1103/PhysRevB.110.184402} {\bibfield  {journal}
  {\bibinfo  {journal} {Phys. Rev. B}\ }\textbf {\bibinfo {volume} {110}},\
  \bibinfo {pages} {184402} (\bibinfo {year} {2024})}\BibitemShut {NoStop}%
\bibitem [{\citenamefont {Sebastian}\ \emph {et~al.}(2021)\citenamefont
  {Sebastian}, \citenamefont {Somesh}, \citenamefont {Nandi}, \citenamefont
  {Ahmed}, \citenamefont {Bag}, \citenamefont {Baenitz}, \citenamefont {Koo},
  \citenamefont {Sichelschmidt}, \citenamefont {Tsirlin}, \citenamefont
  {Furukawa},\ and\ \citenamefont {Nath}}]{PhysRevB.103.064413}%
  \BibitemOpen
  \bibfield  {author} {\bibinfo {author} {\bibfnamefont {S.~J.}\ \bibnamefont
  {Sebastian}}, \bibinfo {author} {\bibfnamefont {K.}~\bibnamefont {Somesh}},
  \bibinfo {author} {\bibfnamefont {M.}~\bibnamefont {Nandi}}, \bibinfo
  {author} {\bibfnamefont {N.}~\bibnamefont {Ahmed}}, \bibinfo {author}
  {\bibfnamefont {P.}~\bibnamefont {Bag}}, \bibinfo {author} {\bibfnamefont
  {M.}~\bibnamefont {Baenitz}}, \bibinfo {author} {\bibfnamefont
  {B.}~\bibnamefont {Koo}}, \bibinfo {author} {\bibfnamefont {J.}~\bibnamefont
  {Sichelschmidt}}, \bibinfo {author} {\bibfnamefont {A.~A.}\ \bibnamefont
  {Tsirlin}}, \bibinfo {author} {\bibfnamefont {Y.}~\bibnamefont {Furukawa}},\
  and\ \bibinfo {author} {\bibfnamefont {R.}~\bibnamefont {Nath}},\ }\bibfield
  {title} {\bibinfo {title} {Quasi-one-dimensional magnetism in the spin-1/2
  antiferromagnet
  \text{Ba}\text{Na$_{2}$}\text{Cu}(\text{V}\text{O$_{4}$)$_{2}$ }},\ }\href
  {https://doi.org/10.1103/PhysRevB.103.064413} {\bibfield  {journal} {\bibinfo
   {journal} {Phys. Rev. B}\ }\textbf {\bibinfo {volume} {103}},\ \bibinfo
  {pages} {064413} (\bibinfo {year} {2021})}\BibitemShut {NoStop}%
\bibitem [{\citenamefont {Khatua}\ \emph {et~al.}(2022)\citenamefont {Khatua},
  \citenamefont {Gomil{\v{s}}ek}, \citenamefont {Orain}, \citenamefont
  {Strydom}, \citenamefont {Jagli{\v{c}}i{\'{c}}}, \citenamefont {Colin},
  \citenamefont {Petit}, \citenamefont {Ozarowski}, \citenamefont
  {Mangin-Thro}, \citenamefont {Sethupathi}, \citenamefont {Rao}, \citenamefont
  {Zorko},\ and\ \citenamefont {Khuntia}}]{Khatua2022}%
  \BibitemOpen
  \bibfield  {author} {\bibinfo {author} {\bibfnamefont {J.}~\bibnamefont
  {Khatua}}, \bibinfo {author} {\bibfnamefont {M.}~\bibnamefont
  {Gomil{\v{s}}ek}}, \bibinfo {author} {\bibfnamefont {J.~C.}\ \bibnamefont
  {Orain}}, \bibinfo {author} {\bibfnamefont {A.~M.}\ \bibnamefont {Strydom}},
  \bibinfo {author} {\bibfnamefont {Z.}~\bibnamefont {Jagli{\v{c}}i{\'{c}}}},
  \bibinfo {author} {\bibfnamefont {C.~V.}\ \bibnamefont {Colin}}, \bibinfo
  {author} {\bibfnamefont {S.}~\bibnamefont {Petit}}, \bibinfo {author}
  {\bibfnamefont {A.}~\bibnamefont {Ozarowski}}, \bibinfo {author}
  {\bibfnamefont {L.}~\bibnamefont {Mangin-Thro}}, \bibinfo {author}
  {\bibfnamefont {K.}~\bibnamefont {Sethupathi}}, \bibinfo {author}
  {\bibfnamefont {M.~S.~R.}\ \bibnamefont {Rao}}, \bibinfo {author}
  {\bibfnamefont {A.}~\bibnamefont {Zorko}},\ and\ \bibinfo {author}
  {\bibfnamefont {P.}~\bibnamefont {Khuntia}},\ }\bibfield  {title} {\bibinfo
  {title} {Signature of a randomness-driven spin-liquid state in a frustrated
  magnet},\ }\href {https://doi.org/10.1038/s42005-022-00879-2} {\bibfield
  {journal} {\bibinfo  {journal} {Commun. Phys.}\ }\textbf {\bibinfo {volume}
  {5}},\ \bibinfo {pages} {99} (\bibinfo {year} {2022})}\BibitemShut {NoStop}%
\bibitem [{\citenamefont {Kinyon}\ \emph {et~al.}(2021)\citenamefont {Kinyon},
  \citenamefont {Dalal}, \citenamefont {Clark}, \citenamefont {Zhou},\ and\
  \citenamefont {Choi}}]{PhysRevMaterials.5.054413}%
  \BibitemOpen
  \bibfield  {author} {\bibinfo {author} {\bibfnamefont {J.~S.}\ \bibnamefont
  {Kinyon}}, \bibinfo {author} {\bibfnamefont {N.~S.}\ \bibnamefont {Dalal}},
  \bibinfo {author} {\bibfnamefont {R.~J.}\ \bibnamefont {Clark}}, \bibinfo
  {author} {\bibfnamefont {H.}~\bibnamefont {Zhou}},\ and\ \bibinfo {author}
  {\bibfnamefont {K.~Y.}\ \bibnamefont {Choi}},\ }\bibfield  {title} {\bibinfo
  {title} {Closing the spin gap of
  \text{(NH$_{4}$)$_{x}$}\text{K$_{1-x}$}\text{Cu}\text{Cl$_{3}$} through
  chemical substitution},\ }\href
  {https://doi.org/10.1103/PhysRevMaterials.5.054413} {\bibfield  {journal}
  {\bibinfo  {journal} {Phys. Rev. Mater.}\ }\textbf {\bibinfo {volume} {5}},\
  \bibinfo {pages} {054413} (\bibinfo {year} {2021})}\BibitemShut {NoStop}%
\bibitem [{\citenamefont {Sebastian}\ \emph {et~al.}(2006)\citenamefont
  {Sebastian}, \citenamefont {Harrison}, \citenamefont {Batista}, \citenamefont
  {Balicas}, \citenamefont {Jaime}, \citenamefont {Sharma}, \citenamefont
  {Kawashima},\ and\ \citenamefont {Fisher}}]{Sebastian2006}%
  \BibitemOpen
  \bibfield  {author} {\bibinfo {author} {\bibfnamefont {S.~E.}\ \bibnamefont
  {Sebastian}}, \bibinfo {author} {\bibfnamefont {N.}~\bibnamefont {Harrison}},
  \bibinfo {author} {\bibfnamefont {C.~D.}\ \bibnamefont {Batista}}, \bibinfo
  {author} {\bibfnamefont {L.}~\bibnamefont {Balicas}}, \bibinfo {author}
  {\bibfnamefont {M.}~\bibnamefont {Jaime}}, \bibinfo {author} {\bibfnamefont
  {P.~A.}\ \bibnamefont {Sharma}}, \bibinfo {author} {\bibfnamefont
  {N.}~\bibnamefont {Kawashima}},\ and\ \bibinfo {author} {\bibfnamefont
  {I.~R.}\ \bibnamefont {Fisher}},\ }\bibfield  {title} {\bibinfo {title}
  {Dimensional reduction at a quantum critical point},\ }\href
  {https://doi.org/10.1038/nature04732} {\bibfield  {journal} {\bibinfo
  {journal} {Nature}\ }\textbf {\bibinfo {volume} {441}},\ \bibinfo {pages}
  {617} (\bibinfo {year} {2006})}\BibitemShut {NoStop}%
\bibitem [{\citenamefont {Benner}\ \emph {et~al.}(1983)\citenamefont {Benner},
  \citenamefont {Brodehl}, \citenamefont {Seitz},\ and\ \citenamefont
  {Wiese}}]{HBenner1983}%
  \BibitemOpen
  \bibfield  {author} {\bibinfo {author} {\bibfnamefont {H.}~\bibnamefont
  {Benner}}, \bibinfo {author} {\bibfnamefont {M.}~\bibnamefont {Brodehl}},
  \bibinfo {author} {\bibfnamefont {H.}~\bibnamefont {Seitz}},\ and\ \bibinfo
  {author} {\bibfnamefont {J.}~\bibnamefont {Wiese}},\ }\bibfield  {title}
  {\bibinfo {title} {Influence of nondiagonal dynamic susceptibility on the
  \text{EPR} signal of \text{H}eisenberg magnets},\ }\href
  {https://doi.org/10.1088/0022-3719/16/31/015} {\bibfield  {journal} {\bibinfo
   {journal} {J. Phys. C: Solid State Phys.}\ }\textbf {\bibinfo {volume}
  {16}},\ \bibinfo {pages} {6011} (\bibinfo {year} {1983})}\BibitemShut
  {NoStop}%
\bibitem [{\citenamefont {Kubo}\ and\ \citenamefont
  {Tomita}(1954)}]{doi:10.1143/JPSJ.9.888}%
  \BibitemOpen
  \bibfield  {author} {\bibinfo {author} {\bibfnamefont {R.}~\bibnamefont
  {Kubo}}\ and\ \bibinfo {author} {\bibfnamefont {K.}~\bibnamefont {Tomita}},\
  }\bibfield  {title} {\bibinfo {title} {A general theory of magnetic resonance
  absorption},\ }\href {https://doi.org/10.1143/JPSJ.9.888} {\bibfield
  {journal} {\bibinfo  {journal} {J. Phys. Soc. Jpn.}\ }\textbf {\bibinfo
  {volume} {9}},\ \bibinfo {pages} {888} (\bibinfo {year} {1954})}\BibitemShut
  {NoStop}%
\bibitem [{\citenamefont {Zorko}\ \emph {et~al.}(2009)\citenamefont {Zorko},
  \citenamefont {Nellutla}, \citenamefont {van Tol}, \citenamefont {Brunel},
  \citenamefont {Bert}, \citenamefont {Duc}, \citenamefont {Trombe},\ and\
  \citenamefont {Mendels}}]{AZorko}%
  \BibitemOpen
  \bibfield  {author} {\bibinfo {author} {\bibfnamefont {A.}~\bibnamefont
  {Zorko}}, \bibinfo {author} {\bibfnamefont {S.}~\bibnamefont {Nellutla}},
  \bibinfo {author} {\bibfnamefont {J.}~\bibnamefont {van Tol}}, \bibinfo
  {author} {\bibfnamefont {L.~C.}\ \bibnamefont {Brunel}}, \bibinfo {author}
  {\bibfnamefont {F.}~\bibnamefont {Bert}}, \bibinfo {author} {\bibfnamefont
  {F.}~\bibnamefont {Duc}}, \bibinfo {author} {\bibfnamefont {J.-C.}\
  \bibnamefont {Trombe}},\ and\ \bibinfo {author} {\bibfnamefont
  {P.}~\bibnamefont {Mendels}},\ }\bibfield  {title} {\bibinfo {title}
  {Electron spin resonance investigation of the spin-1/2 kagomé
  antiferromagnet \text{Zn}\text{Cu$_{3}$}\text{(OH)$_{6}$}\text{Cl$_{2}$}},\
  }\href {https://doi.org/10.1088/1742-6596/145/1/012014} {\bibfield  {journal}
  {\bibinfo  {journal} {J. Phys. Conf. Ser.}\ }\textbf {\bibinfo {volume}
  {145}},\ \bibinfo {pages} {012014} (\bibinfo {year} {2009})}\BibitemShut
  {NoStop}%
\bibitem [{\citenamefont {Oshikawa}\ and\ \citenamefont
  {Affleck}(2002)}]{PhysRevB.65.134410}%
  \BibitemOpen
  \bibfield  {author} {\bibinfo {author} {\bibfnamefont {M.}~\bibnamefont
  {Oshikawa}}\ and\ \bibinfo {author} {\bibfnamefont {I.}~\bibnamefont
  {Affleck}},\ }\bibfield  {title} {\bibinfo {title} {Electron spin resonance
  in $s=\frac{1}{2}$ antiferromagnetic chains},\ }\href
  {https://doi.org/10.1103/PhysRevB.65.134410} {\bibfield  {journal} {\bibinfo
  {journal} {Phys. Rev. B}\ }\textbf {\bibinfo {volume} {65}},\ \bibinfo
  {pages} {134410} (\bibinfo {year} {2002})}\BibitemShut {NoStop}%
\bibitem [{\citenamefont {Bonner}\ and\ \citenamefont
  {Fisher}(1964)}]{PhysRev.135.A640}%
  \BibitemOpen
  \bibfield  {author} {\bibinfo {author} {\bibfnamefont {J.~C.}\ \bibnamefont
  {Bonner}}\ and\ \bibinfo {author} {\bibfnamefont {M.~E.}\ \bibnamefont
  {Fisher}},\ }\bibfield  {title} {\bibinfo {title} {Linear magnetic chains
  with anisotropic coupling},\ }\href
  {https://doi.org/10.1103/PhysRev.135.A640} {\bibfield  {journal} {\bibinfo
  {journal} {Phys. Rev.}\ }\textbf {\bibinfo {volume} {135}},\ \bibinfo {pages}
  {A640} (\bibinfo {year} {1964})}\BibitemShut {NoStop}%
\bibitem [{\citenamefont {Fujisawa}\ \emph {et~al.}(2009)\citenamefont
  {Fujisawa}, \citenamefont {Shiraki}, \citenamefont {Okubo}, \citenamefont
  {Ohta}, \citenamefont {Yoshida}, \citenamefont {Tanaka},\ and\ \citenamefont
  {Sakai}}]{PhysRevB.80.012408}%
  \BibitemOpen
  \bibfield  {author} {\bibinfo {author} {\bibfnamefont {M.}~\bibnamefont
  {Fujisawa}}, \bibinfo {author} {\bibfnamefont {K.}~\bibnamefont {Shiraki}},
  \bibinfo {author} {\bibfnamefont {S.}~\bibnamefont {Okubo}}, \bibinfo
  {author} {\bibfnamefont {H.}~\bibnamefont {Ohta}}, \bibinfo {author}
  {\bibfnamefont {M.}~\bibnamefont {Yoshida}}, \bibinfo {author} {\bibfnamefont
  {H.}~\bibnamefont {Tanaka}},\ and\ \bibinfo {author} {\bibfnamefont
  {T.}~\bibnamefont {Sakai}},\ }\bibfield  {title} {\bibinfo {title}
  {Dzyaloshinsky-moriya interaction in the $s=\frac{1}{2}$
  quasi-one-dimensional antiferromagnet \text{Cu$_{2}$}\text{Cl$_{4}$} $\cdot$
  \text{H$_{8}$}\text{C$_{4}$\text{S}\text{O$_{2}$}} as determined via
  high-frequency esr},\ }\href {https://doi.org/10.1103/PhysRevB.80.012408}
  {\bibfield  {journal} {\bibinfo  {journal} {Phys. Rev. B}\ }\textbf {\bibinfo
  {volume} {80}},\ \bibinfo {pages} {012408} (\bibinfo {year}
  {2009})}\BibitemShut {NoStop}%
\bibitem [{\citenamefont {Choi}\ \emph {et~al.}(2014)\citenamefont {Choi},
  \citenamefont {Choi}, \citenamefont {Lemmens}, \citenamefont {van Tol},\ and\
  \citenamefont {Berger}}]{Choi_2014}%
  \BibitemOpen
  \bibfield  {author} {\bibinfo {author} {\bibfnamefont {K.-Y.}\ \bibnamefont
  {Choi}}, \bibinfo {author} {\bibfnamefont {I.~H.}\ \bibnamefont {Choi}},
  \bibinfo {author} {\bibfnamefont {P.}~\bibnamefont {Lemmens}}, \bibinfo
  {author} {\bibfnamefont {J.}~\bibnamefont {van Tol}},\ and\ \bibinfo {author}
  {\bibfnamefont {H.}~\bibnamefont {Berger}},\ }\bibfield  {title} {\bibinfo
  {title} {Magnetic, structural, and electronic properties of the multiferroic
  compound \text{Fe}\text{Te$_{2}$}\text{O$_{5}$}\text{Br} with geometrical
  frustration},\ }\href {https://doi.org/10.1088/0953-8984/26/8/086001}
  {\bibfield  {journal} {\bibinfo  {journal} {J. Condens. Matter Phys.}\
  }\textbf {\bibinfo {volume} {26}},\ \bibinfo {pages} {086001} (\bibinfo
  {year} {2014})}\BibitemShut {NoStop}%
\bibitem [{\citenamefont {Chabre}\ \emph {et~al.}(2005)\citenamefont {Chabre},
  \citenamefont {Ghorayeb}, \citenamefont {Millet}, \citenamefont
  {Pashchenko},\ and\ \citenamefont {Stepanov}}]{PhysRevB.72.012415}%
  \BibitemOpen
  \bibfield  {author} {\bibinfo {author} {\bibfnamefont {F.}~\bibnamefont
  {Chabre}}, \bibinfo {author} {\bibfnamefont {A.~M.}\ \bibnamefont
  {Ghorayeb}}, \bibinfo {author} {\bibfnamefont {P.}~\bibnamefont {Millet}},
  \bibinfo {author} {\bibfnamefont {V.~A.}\ \bibnamefont {Pashchenko}},\ and\
  \bibinfo {author} {\bibfnamefont {A.}~\bibnamefont {Stepanov}},\ }\bibfield
  {title} {\bibinfo {title} {Low-temperature behavior of the esr linewidth in a
  system with a spin gap:
  $\eta$-\text{Na$_{1.286}$}\text{V$_{2}$}\text{O$_{5}$}},\ }\href
  {https://doi.org/10.1103/PhysRevB.72.012415} {\bibfield  {journal} {\bibinfo
  {journal} {Phys. Rev. B}\ }\textbf {\bibinfo {volume} {72}},\ \bibinfo
  {pages} {012415} (\bibinfo {year} {2005})}\BibitemShut {NoStop}%
\bibitem [{\citenamefont {Glamazda}\ \emph {et~al.}(2017)\citenamefont
  {Glamazda}, \citenamefont {Choi}, \citenamefont {Do}, \citenamefont {Lee},
  \citenamefont {Lemmens}, \citenamefont {Ponomaryov}, \citenamefont {Zvyagin},
  \citenamefont {Wosnitza}, \citenamefont {Sari}, \citenamefont {Watanabe},\
  and\ \citenamefont {Choi}}]{PhysRevB.95.184430}%
  \BibitemOpen
  \bibfield  {author} {\bibinfo {author} {\bibfnamefont {A.}~\bibnamefont
  {Glamazda}}, \bibinfo {author} {\bibfnamefont {Y.~S.}\ \bibnamefont {Choi}},
  \bibinfo {author} {\bibfnamefont {S.-H.}\ \bibnamefont {Do}}, \bibinfo
  {author} {\bibfnamefont {S.}~\bibnamefont {Lee}}, \bibinfo {author}
  {\bibfnamefont {P.}~\bibnamefont {Lemmens}}, \bibinfo {author} {\bibfnamefont
  {A.~N.}\ \bibnamefont {Ponomaryov}}, \bibinfo {author} {\bibfnamefont
  {S.~A.}\ \bibnamefont {Zvyagin}}, \bibinfo {author} {\bibfnamefont
  {J.}~\bibnamefont {Wosnitza}}, \bibinfo {author} {\bibfnamefont {D.~P.}\
  \bibnamefont {Sari}}, \bibinfo {author} {\bibfnamefont {I.}~\bibnamefont
  {Watanabe}},\ and\ \bibinfo {author} {\bibfnamefont {K.-Y.}\ \bibnamefont
  {Choi}},\ }\bibfield  {title} {\bibinfo {title} {Quantum criticality in the
  coupled two-leg spin ladder \text{Ba$_{2}$}{CuTe}\text{O$_{6}$}},\ }\href
  {https://doi.org/10.1103/PhysRevB.95.184430} {\bibfield  {journal} {\bibinfo
  {journal} {Phys. Rev. B}\ }\textbf {\bibinfo {volume} {95}},\ \bibinfo
  {pages} {184430} (\bibinfo {year} {2017})}\BibitemShut {NoStop}%
\bibitem [{\citenamefont {Zorko}\ \emph {et~al.}(2004)\citenamefont {Zorko},
  \citenamefont {Arčon}, \citenamefont {Nuttall},\ and\ \citenamefont
  {Lappas}}]{ZORKO2004E699}%
  \BibitemOpen
  \bibfield  {author} {\bibinfo {author} {\bibfnamefont {A.}~\bibnamefont
  {Zorko}}, \bibinfo {author} {\bibfnamefont {D.}~\bibnamefont {Arčon}},
  \bibinfo {author} {\bibfnamefont {C.~J.}\ \bibnamefont {Nuttall}},\ and\
  \bibinfo {author} {\bibfnamefont {A.}~\bibnamefont {Lappas}},\ }\bibfield
  {title} {\bibinfo {title} {X-band esr study of the 2\text{D} spin-gap system
  \text{Sr}\text{Cu$_{2}$}\text{(BO$_{3}$)$_{2}$}},\ }\href
  {https://doi.org/https://doi.org/10.1016/j.jmmm.2003.12.336} {\bibfield
  {journal} {\bibinfo  {journal} {J. Magn. Magn. Mater.}\ }\textbf {\bibinfo
  {volume} {272-276}},\ \bibinfo {pages} {E699} (\bibinfo {year} {2004})},\
  \bibinfo {note} {proceedings of the International Conference on Magnetism
  (ICM 2003)}\BibitemShut {NoStop}%
\bibitem [{\citenamefont {Yoon}\ \emph {et~al.}(2021)\citenamefont {Yoon},
  \citenamefont {Lee}, \citenamefont {Lee}, \citenamefont {Park}, \citenamefont
  {Lee}, \citenamefont {Choi}, \citenamefont {Do}, \citenamefont {Choi},
  \citenamefont {Chen}, \citenamefont {Chou}, \citenamefont {Gorbunov},
  \citenamefont {Oshima}, \citenamefont {Ali}, \citenamefont {Singh},
  \citenamefont {Berlie}, \citenamefont {Watanabe},\ and\ \citenamefont
  {Choi}}]{PhysRevMaterials.5.014411}%
  \BibitemOpen
  \bibfield  {author} {\bibinfo {author} {\bibfnamefont {S.}~\bibnamefont
  {Yoon}}, \bibinfo {author} {\bibfnamefont {W.}~\bibnamefont {Lee}}, \bibinfo
  {author} {\bibfnamefont {S.}~\bibnamefont {Lee}}, \bibinfo {author}
  {\bibfnamefont {J.}~\bibnamefont {Park}}, \bibinfo {author} {\bibfnamefont
  {C.~H.}\ \bibnamefont {Lee}}, \bibinfo {author} {\bibfnamefont {Y.~S.}\
  \bibnamefont {Choi}}, \bibinfo {author} {\bibfnamefont {S.-H.}\ \bibnamefont
  {Do}}, \bibinfo {author} {\bibfnamefont {W.-J.}\ \bibnamefont {Choi}},
  \bibinfo {author} {\bibfnamefont {W.-T.}\ \bibnamefont {Chen}}, \bibinfo
  {author} {\bibfnamefont {F.}~\bibnamefont {Chou}}, \bibinfo {author}
  {\bibfnamefont {D.~I.}\ \bibnamefont {Gorbunov}}, \bibinfo {author}
  {\bibfnamefont {Y.}~\bibnamefont {Oshima}}, \bibinfo {author} {\bibfnamefont
  {A.}~\bibnamefont {Ali}}, \bibinfo {author} {\bibfnamefont {Y.}~\bibnamefont
  {Singh}}, \bibinfo {author} {\bibfnamefont {A.}~\bibnamefont {Berlie}},
  \bibinfo {author} {\bibfnamefont {I.}~\bibnamefont {Watanabe}},\ and\
  \bibinfo {author} {\bibfnamefont {K.-Y.}\ \bibnamefont {Choi}},\ }\bibfield
  {title} {\bibinfo {title} {Quantum disordered state in the
  ${J}_{1}\text{\ensuremath{-}}{J}_{2}$ square-lattice antiferromagnet
  \text{Sr$_{2}$}\text{Cu}\text{Te$_{0.95}$}\text{W$_{0.05}$}\text{O$_{6}$}},\
  }\href {https://doi.org/10.1103/PhysRevMaterials.5.014411} {\bibfield
  {journal} {\bibinfo  {journal} {Phys. Rev. Mater.}\ }\textbf {\bibinfo
  {volume} {5}},\ \bibinfo {pages} {014411} (\bibinfo {year}
  {2021})}\BibitemShut {NoStop}%
\bibitem [{\citenamefont {Brill}\ \emph {et~al.}(1995)\citenamefont {Brill},
  \citenamefont {Boucher}, \citenamefont {Voiron}, \citenamefont {Palme},
  \citenamefont {Lüthi}, \citenamefont {Dhalenne},\ and\ \citenamefont
  {Revcolevschi}}]{BRILL19951683}%
  \BibitemOpen
  \bibfield  {author} {\bibinfo {author} {\bibfnamefont {T.}~\bibnamefont
  {Brill}}, \bibinfo {author} {\bibfnamefont {J.}~\bibnamefont {Boucher}},
  \bibinfo {author} {\bibfnamefont {J.}~\bibnamefont {Voiron}}, \bibinfo
  {author} {\bibfnamefont {W.}~\bibnamefont {Palme}}, \bibinfo {author}
  {\bibfnamefont {B.}~\bibnamefont {Lüthi}}, \bibinfo {author} {\bibfnamefont
  {G.}~\bibnamefont {Dhalenne}},\ and\ \bibinfo {author} {\bibfnamefont
  {A.}~\bibnamefont {Revcolevschi}},\ }\bibfield  {title} {\bibinfo {title}
  {Direct observation of the q = 0 spin-peierls gap in
  \text{Cu}\text{Ge}\text{O$_{3}$} by high-field esr},\ }\href
  {https://doi.org/https://doi.org/10.1016/0304-8853(94)00694-6} {\bibfield
  {journal} {\bibinfo  {journal} {J. Magn. Magn. Mater.}\ }\textbf {\bibinfo
  {volume} {140-144}},\ \bibinfo {pages} {1683} (\bibinfo {year} {1995})},\
  \bibinfo {note} {international Conference on Magnetism}\BibitemShut {NoStop}%
\bibitem [{dat()}]{data}%
  \BibitemOpen
  \href@noop {} {}\bibinfo {note} {The data that support the findings of the
  current study are available from the corresponding author upon reasonable
  request.}\BibitemShut {Stop}%
\end{thebibliography}%
\end{document}